\documentclass[final,5p,times,twocolumn]{elsarticle}

\usepackage[utf8]{inputenc}

\usepackage{amssymb}
\usepackage{amsmath}
\usepackage{soul}
\usepackage{soul,color}
\usepackage[colorlinks=true,linkcolor=blue,citecolor=blue,urlcolor=blue,bookmarks=true]{hyperref}
\usepackage{subcaption}
\usepackage{booktabs}
\usepackage{tabularx}
\newcolumntype{Y}{>{\centering\arraybackslash}X}
\usepackage[version=4]{mhchem}
\usepackage{lineno}
\usepackage[capitalise]{cleveref}
\usepackage{siunitx}
\usepackage{orcidlink}

\journal{Nuclear Physics A}

\begin{document}

\begin{frontmatter}

\title{Commissioning of GEM Detectors for the Super Bigbite Spectrometer} %

\author[UConn,UVA]{A. Rathnayake\,\orcidlink{0000000231151763}\corref{cor1}}
\author[JLab,WM]{E.W. Wertz\,\orcidlink{0000000316515546}}
\author[UVA]{V. Gamage\,\orcidlink{0000000165121960}}
\author[UVA]{J. McMurtry\,\orcidlink{0009-0006-8752-9100}}
\author[JLab]{C.H. Leung\,\orcidlink{0000000179073728}}
\author[UVA]{A. Ahmed\,\orcidlink{0000-0002-4500-8853}\fnref{CUO}}
\author[JLab]{I. Albayrak}
\author[UVA]{V.V.N. Arachchillage\,\orcidlink{0009000610992850}}
\author[UVA,JLab]{X. Bai}
\author[UVA]{J.A. Boyd III\,\orcidlink{0000-0002-5360-150X}}
\author[JLab]{A. Camsonne\,\orcidlink{0000-0003-4333-2614}}
\author[INFN-RM1,ISS]{E. Cisbani\,\orcidlink{0000000267748473}}
\author[UVA]{C.W. Cotton\,\orcidlink{0000000321893285}}
\author[UVA]{B.D. Purijjala}
\author[HU]{S. Dhital\,\orcidlink{000900057145862X}}
\author[UVA]{D. Di\,\orcidlink{0000000183440631}}
\author[WM]{K.T. Evans\,\orcidlink{0000000154988008}}
\author[UConn,WM]{E. Fuchey\,\orcidlink{0000000301006052}}
\author[HU]{T. Gautam}
\author[UVA,JLab]{K. Gnanvo\,\orcidlink{0000-0002-5348-0664}}
\author[JLab]{C. Ghosh\,\orcidlink{0000-0003-4274-9813}}
\author[UVA]{S. Jeffas\,\orcidlink{0000000169851821}}
\author[UVA]{S. Jian}
\author[HU]{M. Kohl\,\orcidlink{0000-0003-4342-2863}}
\author[UVA]{N. Liyanage}
\author[JLab]{B.J. Moffit\,\orcidlink{0000-0002-3800-1629}}
\author[INFN-GE]{P. Musico\,\orcidlink{0000000258424135}}
\author[UVA]{H. Nguyen\,\orcidlink{0009000725253628}}
\author[UVA]{M. Nycz\,\orcidlink{0000-0002-3570-9103}\fnref{JLAB}}
\author[INFN-BA]{R. Perrino\,\orcidlink{0000-0002-5764-7337}}
\author[UConn]{A.J.R. Puckett\,\orcidlink{0000000236397463}}
\author[HU]{D.M. Rathnayake}
\author[JLab]{B. Raydo}
\author[INFN-CT]{L.G. Re}
\author[JLab]{B. Sawatzky\,\orcidlink{0000000256370348}}
\author[HU]{M. Suresh\,\orcidlink{0000-0003-2326-4765}}
\author[JLab]{H. Szumila-Vance\,\orcidlink{0000-0003-2548-4016}}
\author[JLab]{B. Wojtsekhowski\,\orcidlink{0000000221609814}}

\cortext[cor1]{Corresponding author. Email: anuruddha@uconn.edu, adr@jlab.org}
\fntext[CUO]{Present address: Carleton University-Ottawa, Ottawa, ON K1S 5B6, Canada.}
\fntext[JLAB]{Present address: Thomas Jefferson National Accelerator Facility, Newport News, 23603, VA, USA.}
\affiliation[UConn]{
             organization={Department of Physics, University of Connecticut},
             city={Storrs},
             postcode={06269},
             state={CT},
             country={USA}}
\affiliation[UVA]{
             organization={Department of Physics, University of Virginia},
             city={Charlottesville},
             postcode={22904},
             state={VA},
             country={USA}}

\affiliation[JLab]{
            organization={Thomas Jefferson National Accelerator Facility},
            city={Newport News},
            postcode={23603},
            state={VA},
            country={USA}}
\affiliation[WM]{
            organization={William \& Mary},
            city={Williamsburg},
            postcode={23187},
            state={VA},
            country={USA}}
\affiliation[INFN-RM1]{
            organization={Istituto Nazionale di Fisica Nucleare, Sezione di Roma, 00185 Roma},
            country={Italy}}
\affiliation[ISS]{
            organization={Istituto Superiore di Sanit{\noexpand\`a}, 00185 Roma},
            country={Italy}}
\affiliation[HU]{
            organization={Hampton University},
            city={Hampton},
            state={VA},
            postcode={23669},
            country={USA}}

\affiliation[INFN-GE]{
            organization={Istituto Nazionale di Fisica Nucleare, Sezione di Genova, 16146 Genova},
            country={Italy}}
\affiliation[INFN-BA]{
            organization={Istituto Nazionale di Fisica Nucleare, Sezione di Bari, 70125 Bari},
            country={Italy}}
\affiliation[INFN-CT]{
            organization={Istituto Nazionale di Fisica Nucleare, Sezione di Catania, 95123 Catania},
            country={Italy}}

\begin{abstract}
    We report on the commissioning and cosmic characterization of the Gas Electron Multiplier (GEM) trackers for the recently completed first round of the Super Bigbite Spectrometer (SBS) experimental program conducted in Hall A of the Thomas Jefferson National Accelerator Facility. The SBS GEM trackers rank as one of the largest deployments of triple GEM detectors in operation and were critical to the success of the SBS program. Categorized by their readout strip orientation, the 56 individual SBS GEM modules range in active areas from \qtyproduct{60x50}{\cm} (XY) to \qtyproduct{150x40}{\cm} (UV, XW) and were fabricated at the University of Virginia. Large area layers were assembled from four XY GEM modules or a single UV or XW GEM module. The front-end electronics were based on APV25 chips with output sent to custom developed \SI{40}{\MHz} Multi-purpose Digitizers (MPDs). Prior to the installation of the GEM-based tracking systems for SBS experiments, the GEM modules underwent extensive cosmic-ray testing to characterize the performance of each module, including detector stability, efficiency, uniformity, timing performance, and spatial resolution.
\end{abstract}


\end{frontmatter}

\section{Introduction} \label{sec:Introduction} The 12~GeV upgrade of the Continuous Electron Beam Accelerator Facility (CEBAF) \cite{CEBAF12GeV} at Jefferson Lab (JLab) has enabled a new generation of precision experiments. The combination of CEBAF's increased electron beam energy and spin polarization, along with advances in detector technologies, allowed the first phase of the Super Bigbite Spectrometer (SBS) program \cite{wojtsekhowski2026superbigbitespectrometerphysics} to collect data in Jefferson Lab experimental Hall A from 2021 to 2025. The SBS nucleon electromagnetic form factor program at Jefferson Lab extends the squared four-momentum transferred to the scattered nucleon  ($Q^2$) range of $G_M^n$, $G_E^n$, and $G_E^p$ through precision measurements using multiple extraction methods. Data were collected for six experiments: SBS-GMn \cite{GMNProp}, SBS-nTPE \cite{nTPEProp}, SBS-GEn-II \cite{GEnIIProp}, SBS-GEn-RP \cite{GENRPProp}, SBS-$\mathrm{K_{LL}}$ \cite{KLLProp}, and SBS-GEp-V \cite{GEp5Prop}.

\begin{figure}[t]
    \centering
    \includegraphics[width=\linewidth]{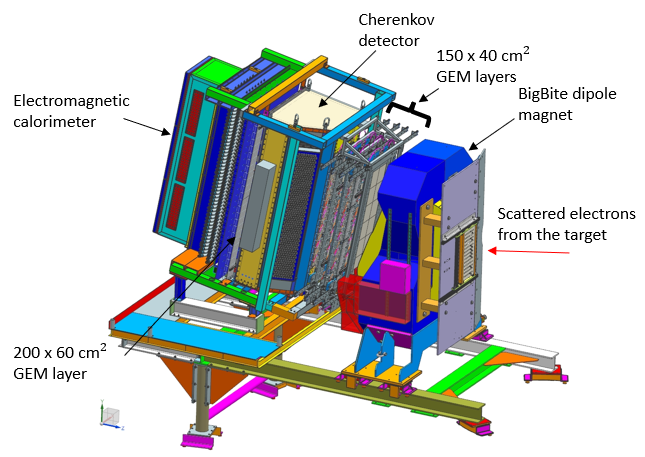}
    \caption{A CAD diagram of the BigBite spectrometer. The main components of the system are labeled, distinguishing the GEM layers installed.}
    \label{fig:bigbite_spectrometer_cad}
\end{figure}

\begin{figure}[t]
    \centering
    \includegraphics[width=\linewidth]{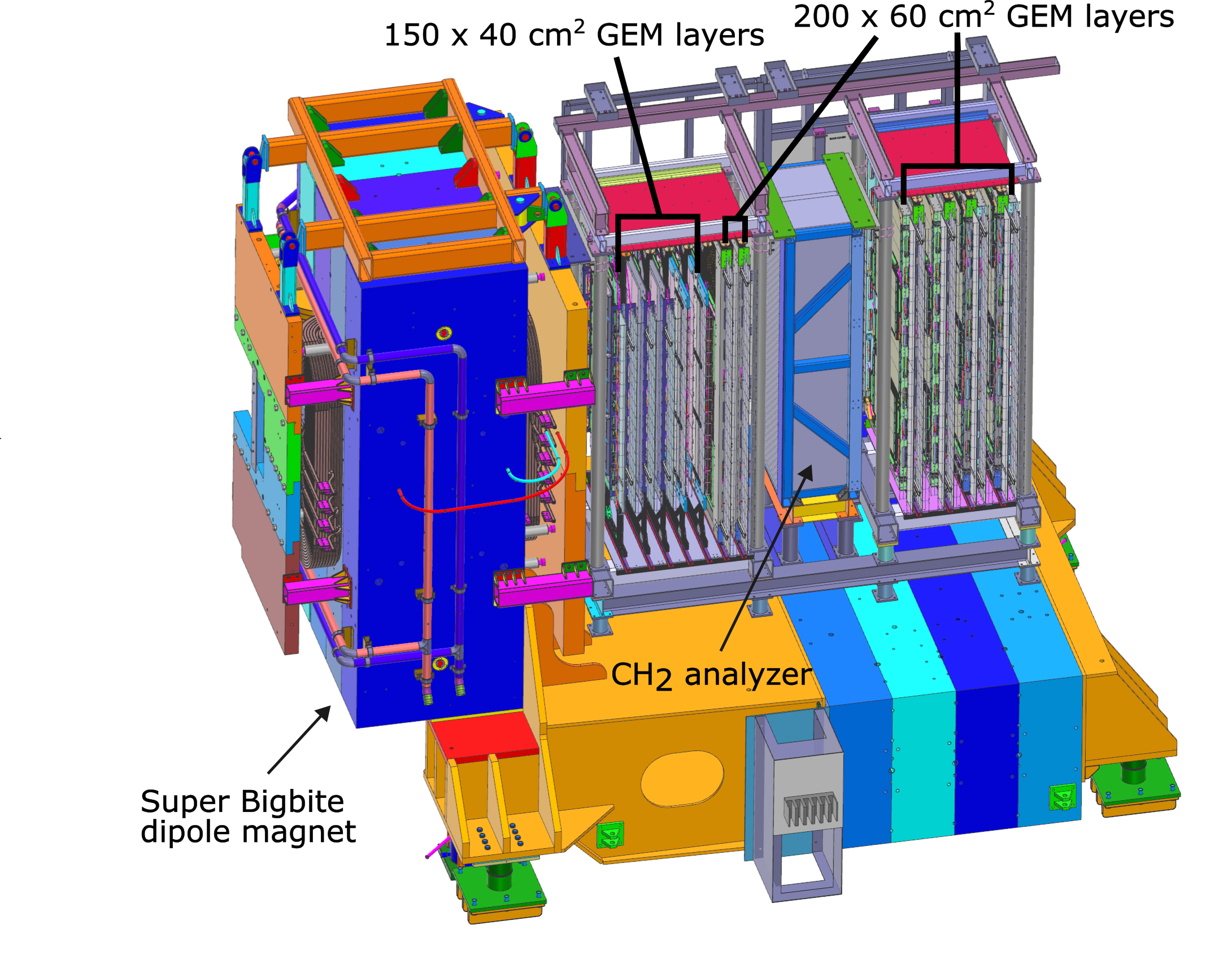}
    \caption{A CAD diagram of the Super Bigbite Spectrometer.}
    \label{fig:gep_hadron_arm_cad}
\end{figure}

The elastic electron-nucleon scattering cross section decreases rapidly as $Q^{-12}$, and as such the experiments in the SBS program, which investigated the very high $Q^{2}$ regime, had to utilize high luminosities to compensate for an elastic scattering probability suppressed by the low cross sections. Consequently, this required detectors capable of operating in high rate environments. The SBS experimental program utilized two medium-acceptance open geometry spectrometers (BigBite and Super Bigbite Spectrometers) which were instrumented with many layers of triple Gas Electron Multiplier (GEM) tracking detectors. The BigBite Spectrometer detected scattered electrons and was composed of a 20~ton large-aperture dipole magnet \cite{DELANGE1998182} and a detector system. The BigBite detector system consisted of five layers of GEM tracking detectors, a gas ring imaging Cherenkov (GRINCH) detector \cite{MSatnik}, a timing hodoscope \cite{RalphThesis}, and preshower and shower calorimeters \cite{BBCalNIM}. \Cref{fig:bigbite_spectrometer_cad} shows a CAD diagram of the BigBite spectrometer. The Super Bigbite Spectrometer \cite{SBSSpecConcept} detected scattered hadrons and was composed of a 100~ton large-aperture dipole magnet, many layers of GEM detectors, and a hadron calorimeter (HCal) \cite{SSeeds}. \Cref{fig:gep_hadron_arm_cad} shows a CAD diagram of the Super Bigbite Spectrometer (excluding HCal), with many GEM layers present.

GEM detectors, known for their ability to operate in high rate environments, were selected for the tracking detector system of the SBS program. For a typical GEM detector, depending on the pitch of the strips and the readout electronics, the spatial resolution can reach an accuracy on the order of \SI{100}{\micro\meter}. One major advantage of the GEM detector technology is the relatively low cost per active area compared to other high-resolution tracking detectors, for example, silicon tracking detectors.

This paper discusses the commissioning of the GEM modules developed for the SBS program, including assembly of the GEM trackers and auxiliary systems, design refinements informed by initial testing, and extensive characterization using cosmic ray testing.

\section{SBS GEM Modules} \label{sec:GEM design and fabrication} The GEM modules for the SBS program were designed and assembled by the Gaseous Detector Group at the University of Virginia using GEM foils produced by CERN's Micro Pattern Technologies workshop~\cite{CERN-MPT}. The SBS GEM modules are based on the proven principle and design of the COMPASS triple-GEM detectors used at CERN~\cite{COMPASS_GEM_performance,GEM2016,COMPASS_GEM_construction}.

Three types of triple GEM modules were fabricated for the SBS program with a naming scheme categorized by their respective readout strip pattern: XY, UV, and XW GEM modules. XY GEM modules have an active area of \qtyproduct{60x50}{\cm} and have orthogonal oriented readout strips in horizontal and vertical directions. The UV GEM modules have an active area of \qtyproduct{150x40}{\cm} and an angle of \ang{60} between the two sets of strips, which are oriented at \ang{\pm30} with respect to the horizontal. The XW GEM modules have the same active area as the UV modules; however, the angle between the two sets of readout strips is \ang{45}, with horizontal strips running parallel to the shorter side of the module. The selection of UV and XW GEM readout strip angles was motivated by the minimization of the readout strip length, as strip capacitance is observed to exhibit a linear relationship with noise in front-end electronics~\cite{BACHMANN2001548}. A total of 50 XY, 4 UV, and 2 XW GEM modules were manufactured for the SBS program.

\begin{figure}
    \centering
    \includegraphics[width=0.95\linewidth]{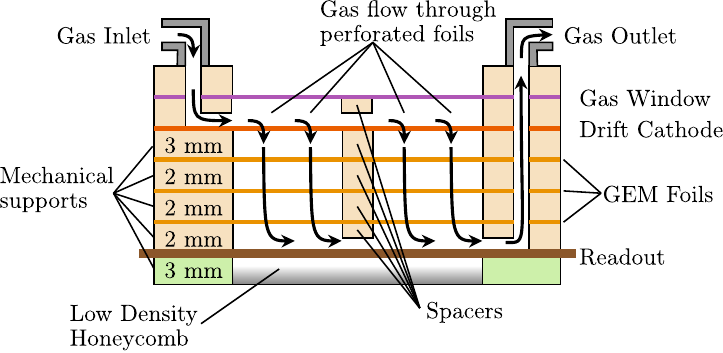}
    \caption{Structural diagram of the SBS XY triple GEM detector design (not to scale). Adapted from Ref. \cite{JBoyd}.} 
    \label{fig:SBS_GEM_design_cross-section}
\end{figure}

Due to the large active area of the SBS GEM modules, additional structural supports were needed. However, concerns over  multiple scattering effects on position resolution also necessitated the minimization of material within the GEM active area. \Cref{fig:SBS_GEM_design_cross-section} illustrates some features that are the result of these requirements: Thin spacer grids attached to the GEM support frames help to ensure a uniform gap between GEM layers. Additionally, all SBS GEM modules feature a low density honeycomb structure beneath the readout board which provides rigid support to the module. UV and XW GEM modules have an additional honeycomb layer on top of the drift cathode. A detailed description of the development, construction and characterization of a prototype SBS GEM module is available in Ref.~\cite{UVASBS_GEM_KG}. A dedicated article providing greater detail on the SBS GEM construction is forthcoming~\cite{SBSGEMConstruct}.

\section{GEM Layer Assembly} \label{sec:Layer assembly} 
\begin{figure}[t]
    \centering
    \includegraphics[width=0.9\linewidth]{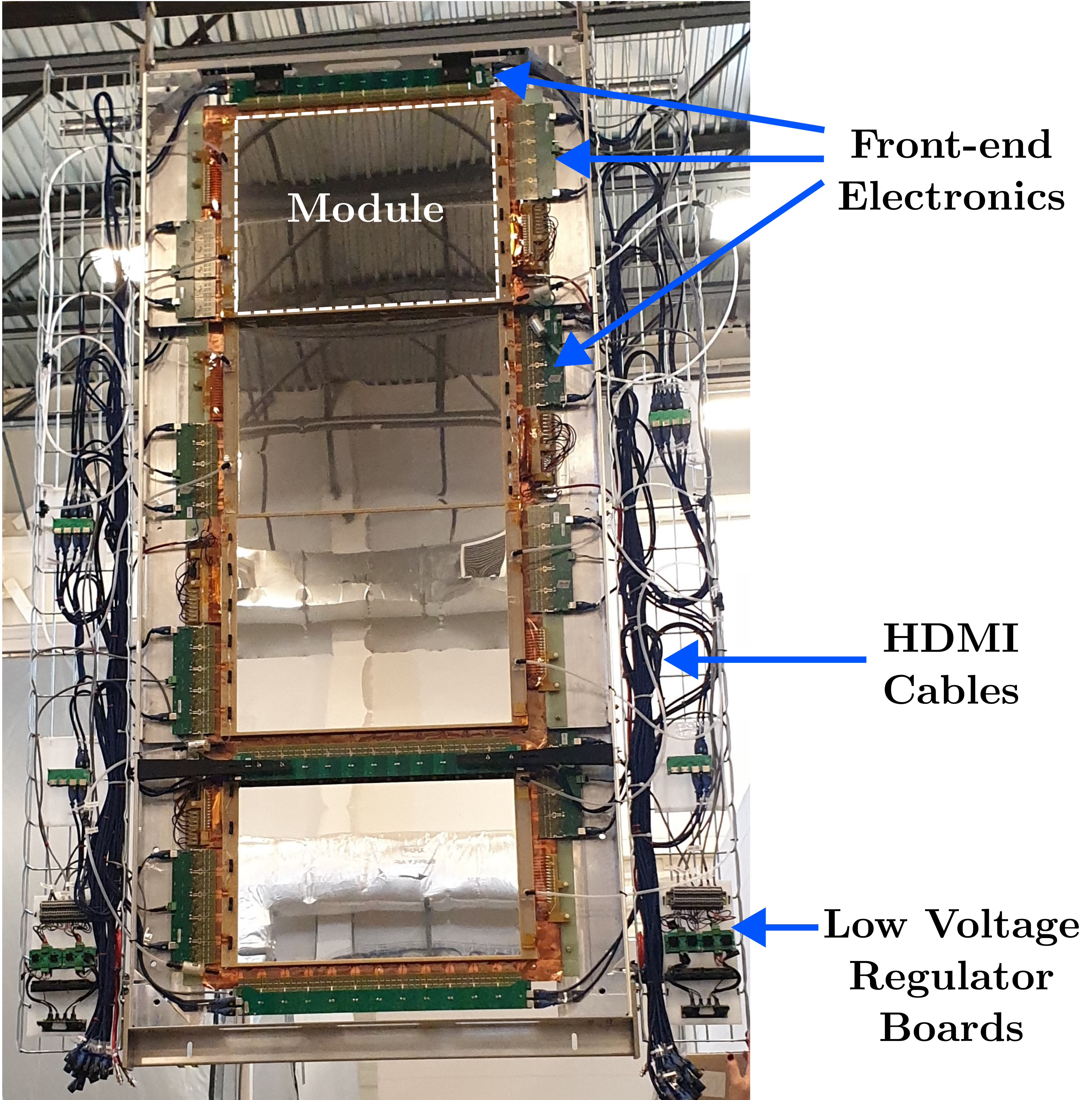}
    \caption{A fully-assembled \qtyproduct{200x60}{\cm} XY GEM layer before RF shielding attachment.}
    \label{fig:XY_GEM_layer_hanging}
\end{figure}

\begin{figure}[t]
    \centering
    \includegraphics[width=0.705\linewidth]{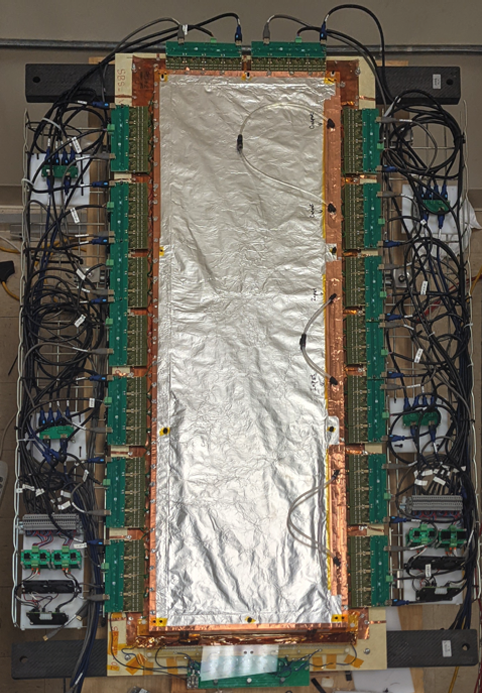}
    \caption{A fully-assembled \qtyproduct{150x40}{\cm} UV GEM layer. A fully-assembled XW GEM layer would be visually similar.}
    \label{fig:UV_GEM_layer}
\end{figure}

After GEM module fabrication and quality-assurance testing were completed, the GEM modules were mounted onto support frames along with the necessary front-end electronics, cables, and auxiliary system infrastructure to form modular GEM tracking layers. XY GEM layers were constructed from four vertically stacked XY GEM modules, forming a composite GEM tracking layer with a total active area of \qtyproduct{200x60}{\cm}. The UV and XW GEM layers had nearly identical GEM tracking layer assemblies, both were single-module GEM layers with an active area of \qtyproduct{150x40}{\cm}. Part of the assembly process for all GEM layers was the installation of high-definition multimedia interface (HDMI) data acquisition cables and auxiliary system components: low-voltage cables, high-voltage cables, and gas tubes. These were installed onto metal cable trays mounted on both sides of the layer support frames. A description of the infrastructures and components of the GEM auxiliary system is presented in \cref{sec:Aux_Syst}. \Cref{fig:XY_GEM_layer_hanging,fig:UV_GEM_layer} show fully-assembled XY (without electromagnetic shielding, described in \cref{sec:GEM shielding}) and UV GEM layers (with electromagnetic shielding). In both \cref{fig:XY_GEM_layer_hanging,fig:UV_GEM_layer}, the APV25 front-end readout cards~\cite{APV25,APVmanual} are shown attached to the readout strips of individual GEM modules, with HDMI cables to transmit signals to additional Data Acquisition (DAQ) electronics. The readout scheme and DAQ electronics are discussed in \cref{sec:Readout scheme}. Each XY, UV, and XW GEM layer is outfitted with 88, 60, and 58 APV25 cards, respectively.

A total of 11 XY, 4 UV and 2 XW GEM layers were produced for the entire SBS experimental program. Typically, the comparatively smaller UV and XW GEM layers were reserved for the front-tracker layers of the spectrometer, while the larger XY GEM layers were reserved for the back-tracker layers, as shown in \cref{fig:bigbite_spectrometer_cad,fig:gep_hadron_arm_cad}, to maintain solid angle acceptance of the spectrometer.

\section{GEM Readout and Data Acquisition} \label{sec:Readout scheme}\subsection{Readout Electronics}
\begin{figure}[ht!]
    \centering
    \includegraphics[width=0.95\linewidth]{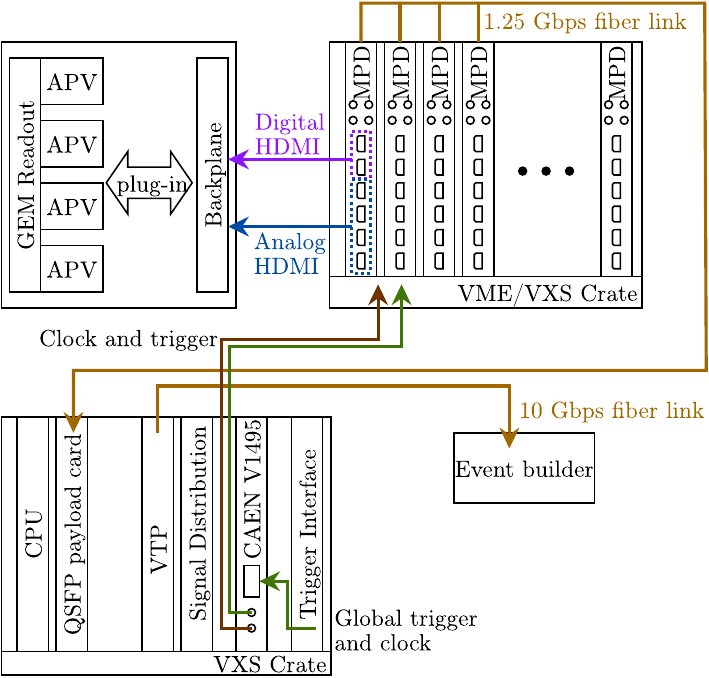}
    \caption{A simplified block diagram of the GEM DAQ system.} 
    \label{fig:GEM_DAQ_Simplified}
\end{figure}

The SBS DAQ uses the CEBAF Online Data Acquisition (CODA) software system developed at JLab~\cite{CODA}. \Cref{fig:GEM_DAQ_Simplified} shows a simplified diagram of the main DAQ components related to the GEM modules used during the SBS program. The APV25~\cite{APV25,APVmanual}-based front-end cards are connected to the detector readout strips, with each card reading out 128 physical strips. The charge collected by the detector strips is continuously sampled by the APV25 chips at a frequency of \SI{41.666}{MHz}, provided by an external clock signal to the APV25.

The APV25 front-end cards are connected to Multi-Purpose Digitizer (MPD) modules via HDMI cables. The MPDs were developed at the Istituto Nazionale di Fisica Nucleare (INFN) for the SBS program~\cite{GEMMPD2013429}. Upon an accepted trigger, the MPD commands the APV25 to release an analog readout frame, which the MPD then digitizes. The MPD hosts up to 16 free running ADCs, which continuously sample the incoming analog signals at the system clock rate, independent of whether a trigger has been accepted. 
The synchronous samples are processed by the on-board Field Programmable Gate Array (FPGA), to identify the APV frames and extract the analog data. The MPD also handles the configuration of the connected APVs, driving the $\mathrm{I^2C}$ (Inter-Integrated Circuit~\cite{I2CPatent}) communication protocol, a two-wire serial interface commonly used for communication between integrated circuits, and providing the required control signals. To ensure signal integrity and noise control, HDMI cable lengths between the APV25 and MPDs were limited to 10--20~m.

The digitized signals are then transmitted to the VXS Trigger Processor (VTP)~\cite{VTP} over optical fibers. The VTP is responsible for packaging the data in CODA format~\cite{CODA} and transferring the data to the DAQ host. During production, the VTP is also used to perform zero suppression online to maximize the overall DAQ bandwidth, as detailed in \cref{sec:Readout_and_OnlineProcessing}.

\subsection{Signal Readout and Online Data Processing}\label{sec:Readout_and_OnlineProcessing}

\begin{figure}
    \centering
\begin{subfigure}{0.95\linewidth}
    \includegraphics[width=\linewidth]{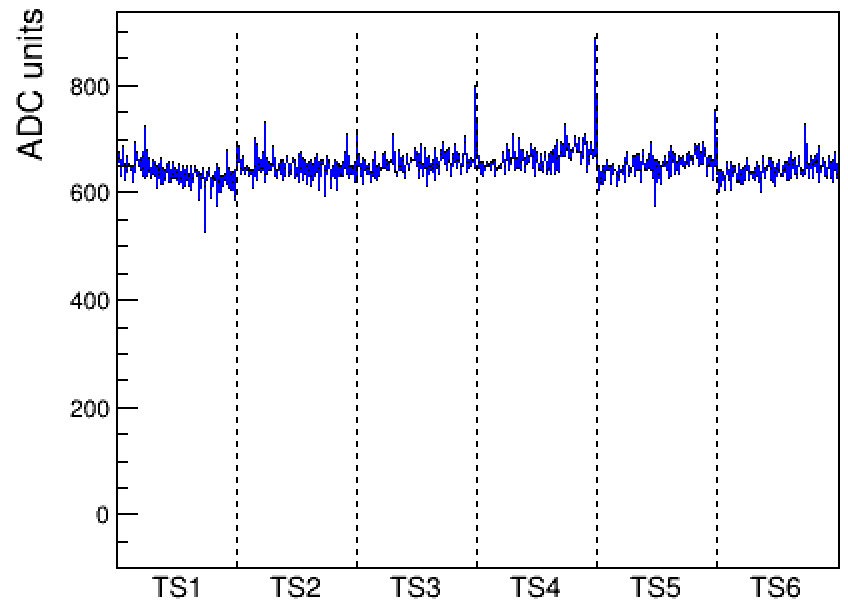}
    \caption{}
    \label{fig:raw_apv_frame_pedestal}
\end{subfigure}
\begin{subfigure}{0.95\linewidth}
    \includegraphics[width=\linewidth]{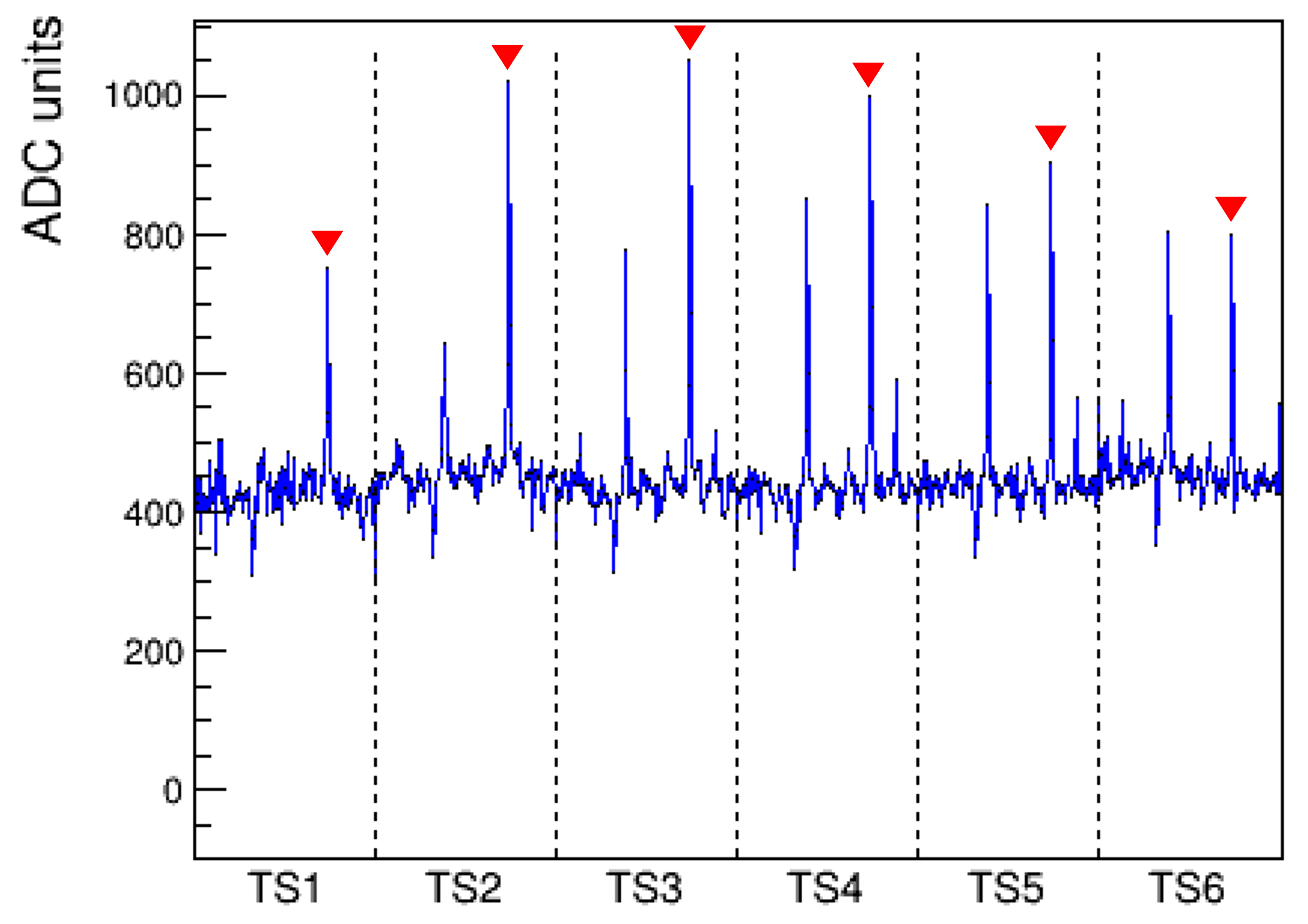}
    \caption{}
    \label{fig:raw_apv_frame_signal}
\end{subfigure}
\caption{Examples of digitized signals for one event of a single APV25 card when all 128 channels are acquired simultaneously. There are 6 time samples per event. Each time sample is separated by \SI{24}{\ns}. Channels 1-128 are ordered sequentially for each consecutive time sample (TS1-TS6), delineated by vertical dashed lines in each plot. The bottom panel shows an example with signals due to an ionization event within the GEM module.}
\end{figure}

For a triggered event, all 128 channels of an APV25 card are sampled multiple times to capture the time progression of the amplified and shaped analog signal pulse. Two examples of digitized data from an APV25 card are presented in \cref{fig:raw_apv_frame_pedestal,fig:raw_apv_frame_signal}. These frames represent the direct output of the MPD following digitization of the analog APV25 signals. For the SBS program, the MPD firmware was programmed to read data 6 times per triggered event, with samples spaced out by \SI{24}{\ns} (as dictated by the external \SI{41.666}{\MHz} clock signal), making the entire readout window \SI{144}{\ns} long. Each plot in \cref{fig:raw_apv_frame_pedestal,fig:raw_apv_frame_signal} shows the channel amplitude in ADC units on the vertical axis, and the 128 channels of the APV25 over the six time samples are shown on the horizontal axis. The 5 long vertical dashed-lines present in both figures are artificially implemented to denote the boundary between time samples.

In \cref{fig:raw_apv_frame_pedestal} no signal resulting from GEM ionization amplification is present in the data-frame; when this data was collected, the high voltage of the GEM detector was off. Data collected without ionization amplification reflect intrinsic electronic baselines and are referred to as pedestal data. The pedestal data are used to quantify the baseline and electronic noise of individual GEM module channels, a requirement before real signals can be extracted. The raw data-frame presented in \cref{fig:raw_apv_frame_signal} represents a case where good ionization signals were detected on multiple channels of the APV25 card. The red downward triangles indicate the channel with the strongest pulse. The shape of the data is consistent with the effect that the pulse shaper circuit on the APV25 chip has on the charge collected from the GEM strips as is modeled by
\begin{equation}\label{PulseShape}
    v(t) = C\frac{t}{T_p} e^{-t/T_p},
\end{equation}
where $T_p=\SI{56}{\ns}$ is the peaking time of the CR-RC pulse, and $C$ is a normalization constant~\cite{APV25}.

\subsubsection{Common-Mode Noise Subtraction and Zero Suppression}

\begin{figure}
    \centering
    \includegraphics[width=1\linewidth]{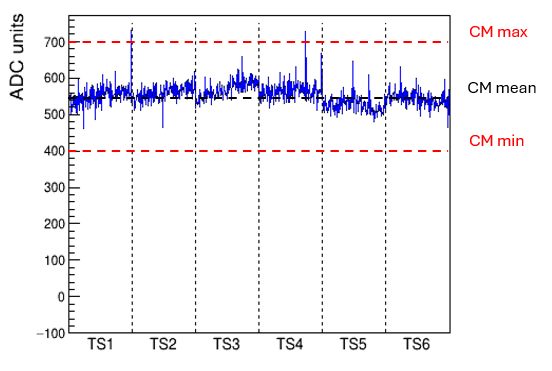}
    \caption{Visualization of a typical CM min and CM max region of an APV25 card, considered for the CM calculation within the VTP.}
    \label{fig:CM_Max_and_Min}
\end{figure}

For a single APV25 card, all 128 channels typically fluctuate by a common offset on a time sample-by-sample and event-by-event basis. This collective channel shift is referred to as the common mode (CM) noise. An example of a raw data-frame where the CM has significantly shifted between the time samples is presented in \cref{fig:CM_Max_and_Min}, and the value of the CM mean for all time samples is represented by the dashed black line. In order to properly extract an amplified signal caused by ionizing particles in the GEM module, the CM must be corrected for each time sample per event for every APV25 card.

To characterize the unique electronic noise of a single APV25 channel, two key parameters are considered: pedestal offset, $\mu_{\rm ped}$, and pedestal rms, $\sigma_{\rm ped}$. For a specific readout channel of an APV25 card, empirical observations show that individual signals remain at a constant offset relative to the CM noise level, exhibiting a narrow distribution around this baseline value. Pedestal distributions for all the GEM detector channels are determined by taking at least several thousand events of pedestal data, done by setting the GEM module high voltages to zero to ensure no ionization events occur during the pedestal data collection.  The pedestal distributions are formed for each of the APV25 channels by taking the average value of the CM-subtracted signals of the six time samples. The mean of this distribution is interpreted as the offset of the pedestal, $\mu_{\rm ped}$, and the noise is quantified by the rms, $\sigma_{\rm ped}$, of the distribution. 

 An algorithm is programmed into the VTP that calculates the CM in real time during data collection. In the case similar to the data frame presented in \cref{fig:raw_apv_frame_pedestal}, where no strips in the GEM module detect a charged particle, the CM calculation for a single time sample is straightforward and would simply be the arithmetic mean of the 128 channels of a given APV25 card. However, for a consideration similar to the data frame shown in \cref{fig:raw_apv_frame_signal}, where some strips have detected the presence of an ionization event in the GEM module, the CM calculation is nontrivial in that the channels corresponding to the fired strips should not bias the CM calculation. A lightweight and memory-efficient CM algorithm~\cite{Danning} was utilized, within the availability of resources in the VTP. The method takes the average of all channels that have ADC units between a pre-determined range of CM values, as determined by analyzing pedestal data for a given APV25 card (see \cref{fig:CM_Max_and_Min} for a representation of this CM range, indicated by red dashed lines). The channels that have ADC values $\sqrt{6}f\sigma_{\rm ped}$ higher or lower than the previously calculated average are removed. Here, $\sigma_{\rm ped}$: pedestal rms of the average of the six time samples, and $f$: an appropriately chosen multiplicative factor, which was 3 the large majority of the time. This process is done iteratively; after each iteration, the upper and lower CM limits are adjusted using the newly calculated CM and $\pm\sqrt{6}f\sigma_{\rm ped}$, thereby progressively excluding the channels containing detector signals. After three such iterations, the resulting average is taken as the CM. 
 
\begin{figure}[t]
    \centering
    \includegraphics[width=0.9\linewidth]{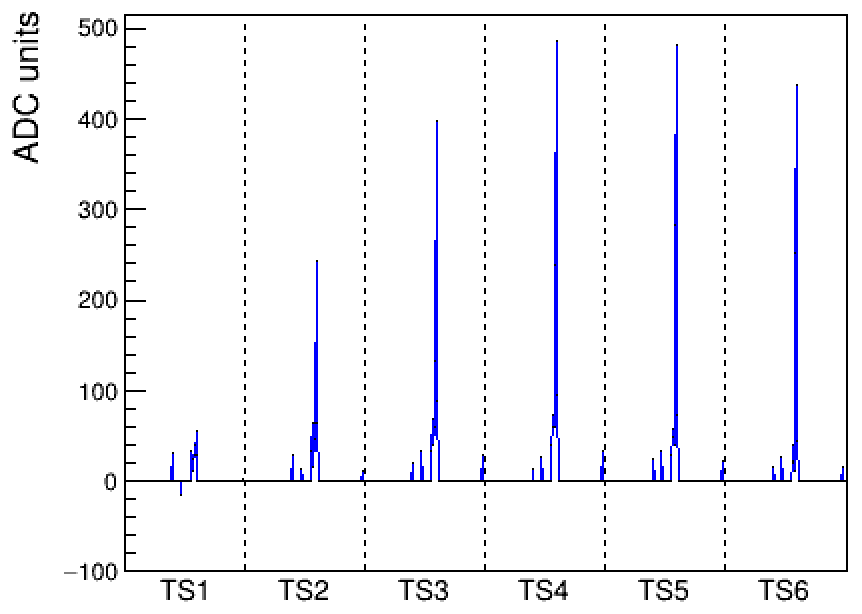}
    \caption{An APV25 frame after online processing.}
    \label{fig:Post online APV frame}
\end{figure}

Once the CM is estimated and subtracted, the true signal strength of that given strip can be determined. Zero-suppression of the strips with ``insignificant'' signal strength was performed by comparing the CM subtracted six time sample average of the given strip with 3$\times\sigma_{\rm ped}$. \Cref{fig:Post online APV frame} depicts an APV25 data frame after online processing - including pedestal subtraction, CM correction, and zero suppression - is complete.

\section{Auxiliary Systems} \label{sec:Aux_Syst} \subsection{High-Voltage Power Supply System}

\subsubsection{Resistive Dividers}

The initial SBS GEM high-voltage (HV) system used a passive resistive chain divider based on the COMPASS triple GEM design~\cite{COMPASS_GEM_construction,UVASBS_GEM_KG}. In this scheme, a single channel was used to supply high voltage to the module, while the individual voltages across the induction gap, transfer gaps, and GEM foils were set by the resistor chain. This configuration worked reliably for most of the GEM modules during early commissioning. However, a few XY GEM modules exhibited unstable behavior, with audible foil vibrations followed by tripping in the high-voltage power supply due to current draw exceeding the nominal range. This mechanical vibration could be attributed to oscillations between imperfectly stretched GEM/drift foils and strong electrostatic interaction between neighboring foils. This was mitigated by reducing the drift, transfer, and induction gap fields by 15$\%$, while keeping the voltage difference across the GEM foils unchanged.

\begin{figure}[ht!]
    \centering
    \includegraphics[width=0.50\textwidth]{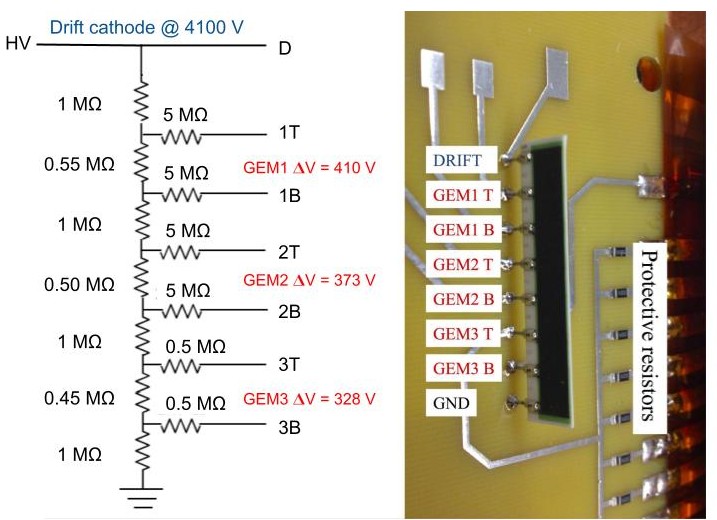}
    \caption{Initial resistive-chain high-voltage divider, reproduced from Ref.~\cite{UVASBS_GEM_KG} 
    }
    \label{Figure:Resistivedivider_0}
\end{figure}

\begin{figure}[ht!]
    \centering
    \includegraphics[width=0.32\textwidth]{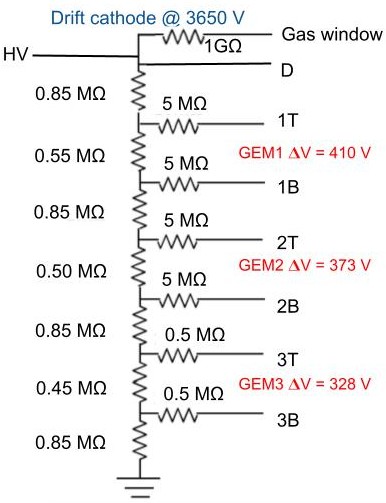}
    \caption{Modified SBS GEM high-voltage divider. The reduced gap-resistor values lower the drift, transfer, and induction fields while preserving the nominal GEM foil voltages
    }
    \label{Figure:Resistivedivider_1}
\end{figure}

This modification was implemented by replacing the original \SI{1}{\Mohm} (\cref{Figure:Resistivedivider_0}) gap resistors with \SI{0.85}{\Mohm}, forming the SBS GEM HV divider (\cref{Figure:Resistivedivider_1}). Subsequent SBS GEM modules were built directly with this modified divider configuration. The XY and UV/XW modules used two different protective resistor configurations. For the XY modules, protective resistors were connected only to the top electrodes of the GEM foils. Whereas for the UV/XW modules, protective resistors were connected to both the top and bottom electrodes. For the XY modules, the protective-resistor values were \SI{10}{\Mohm}, \SI{10}{\Mohm}, and \SI{1}{\Mohm} for GEM1, GEM2, and GEM3, respectively. For the UV/XW modules, the corresponding top- and bottom-electrode resistor values were \SI{5}{\Mohm} and \SI{5}{\Mohm} for GEM1, \SI{5}{\Mohm} and \SI{5}{\Mohm} for GEM2, and \SI{0.5}{\Mohm} and \SI{0.5}{\Mohm} for GEM3. \Cref{Figure:Resistivedivider_0,Figure:Resistivedivider_1,fig:divider_and_individual_scheme} illustrate the UV/XW protective resistor configuration.

\subsubsection{Individual Channel Power Supply}

During high-luminosity operation, the passive high-voltage divider was found to limit the detector gain. As the particle flux increases, the higher hit rate leads to increased ionization and charge deposition in the detector, resulting in a larger current drawn through the high-voltage divider. Current draw is proportional to both the number of ionizations and the detector gain and is expected to increase linearly with particle flux if the gain remains constant. However, during the first experiments of the SBS program, a clear deviation from this expected linear behavior was observed at high rates, indicating a loss of detector gain.

\begin{figure}[!htbp]
    \centering
    \includegraphics[width=0.95\linewidth]{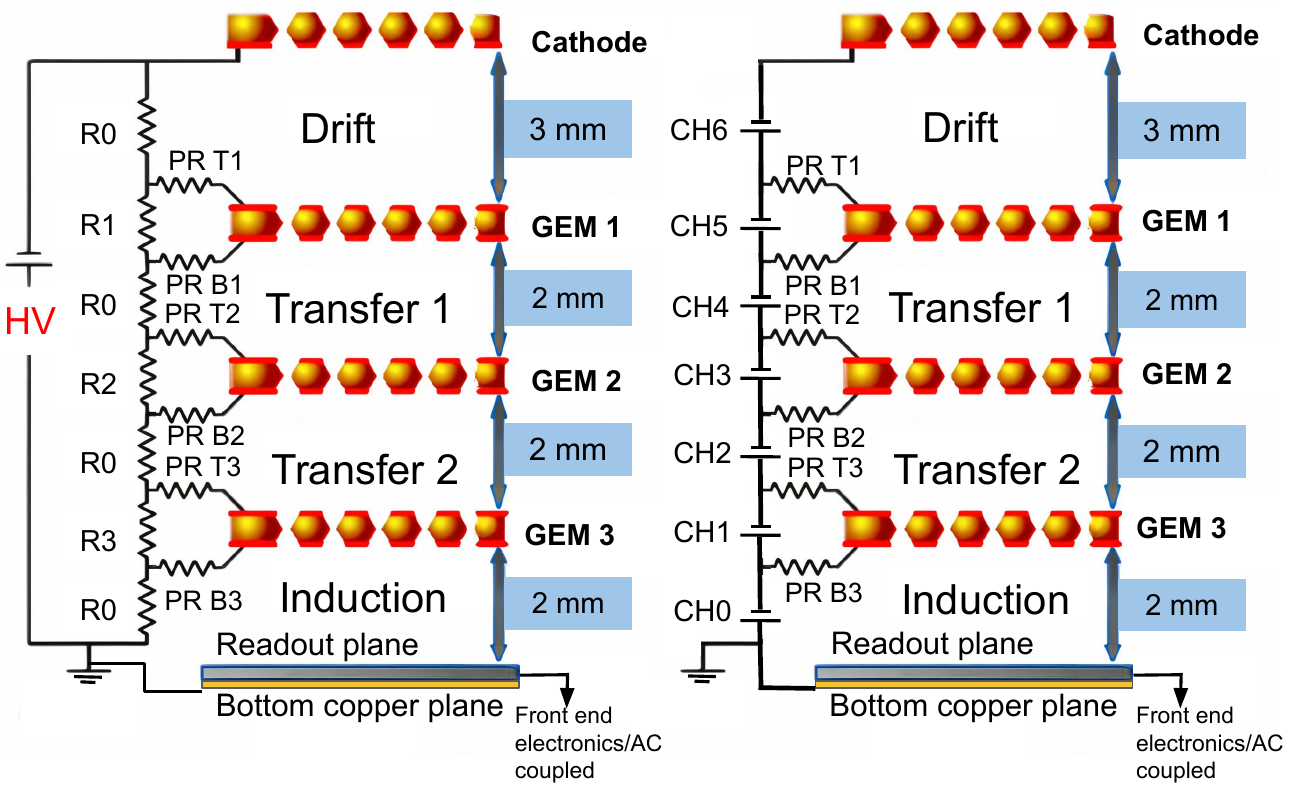}
    \caption{Circuit diagrams for the GEM high-voltage distribution systems used during the SBS experimental program. Left: Passive resistive divider. Right: Active divider, using an individual channel power supply.}
    \label{fig:divider_and_individual_scheme}
\end{figure}

\begin{figure}[!htbp]
    \centering
    \includegraphics[width=0.95\linewidth]{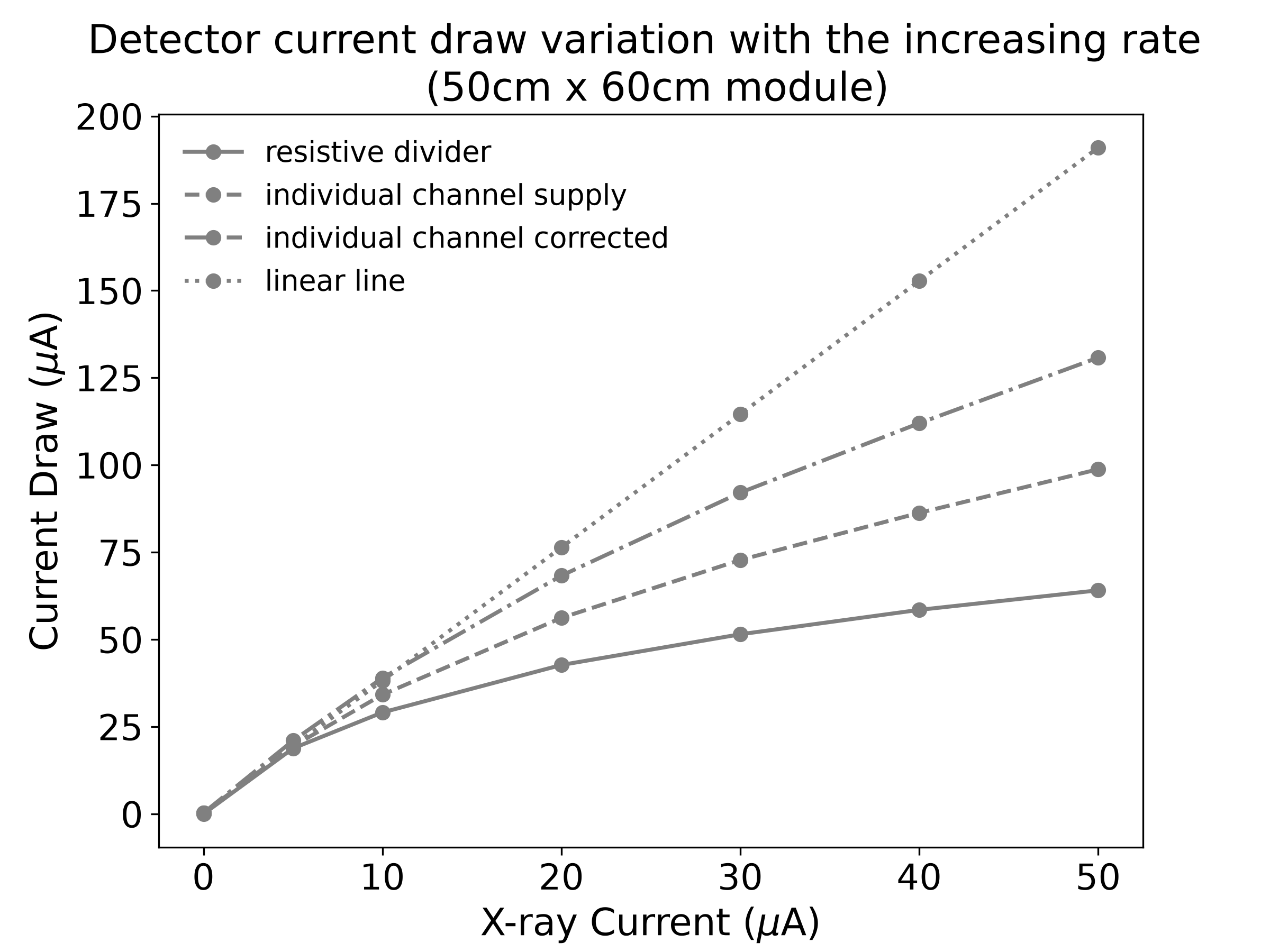}
    \caption{Comparison of current draw as a function of X-ray current for a typical SBS XY GEM module, using different high-voltage system configurations. The current draw using the resistive divider (solid line) deviates from the linear relation (dotted line), indicating a reduction in gain. Using individual channel supply (dashed line) partially mitigate the gain drop. It is further improved by applying corrections for the voltage drop across the current limiting resistors (dotted-dashed line).}
    \label{fig:Excess_current_50x60}
\end{figure}

\begin{figure}[htbp]
    \centering
    \includegraphics[width=1\linewidth]{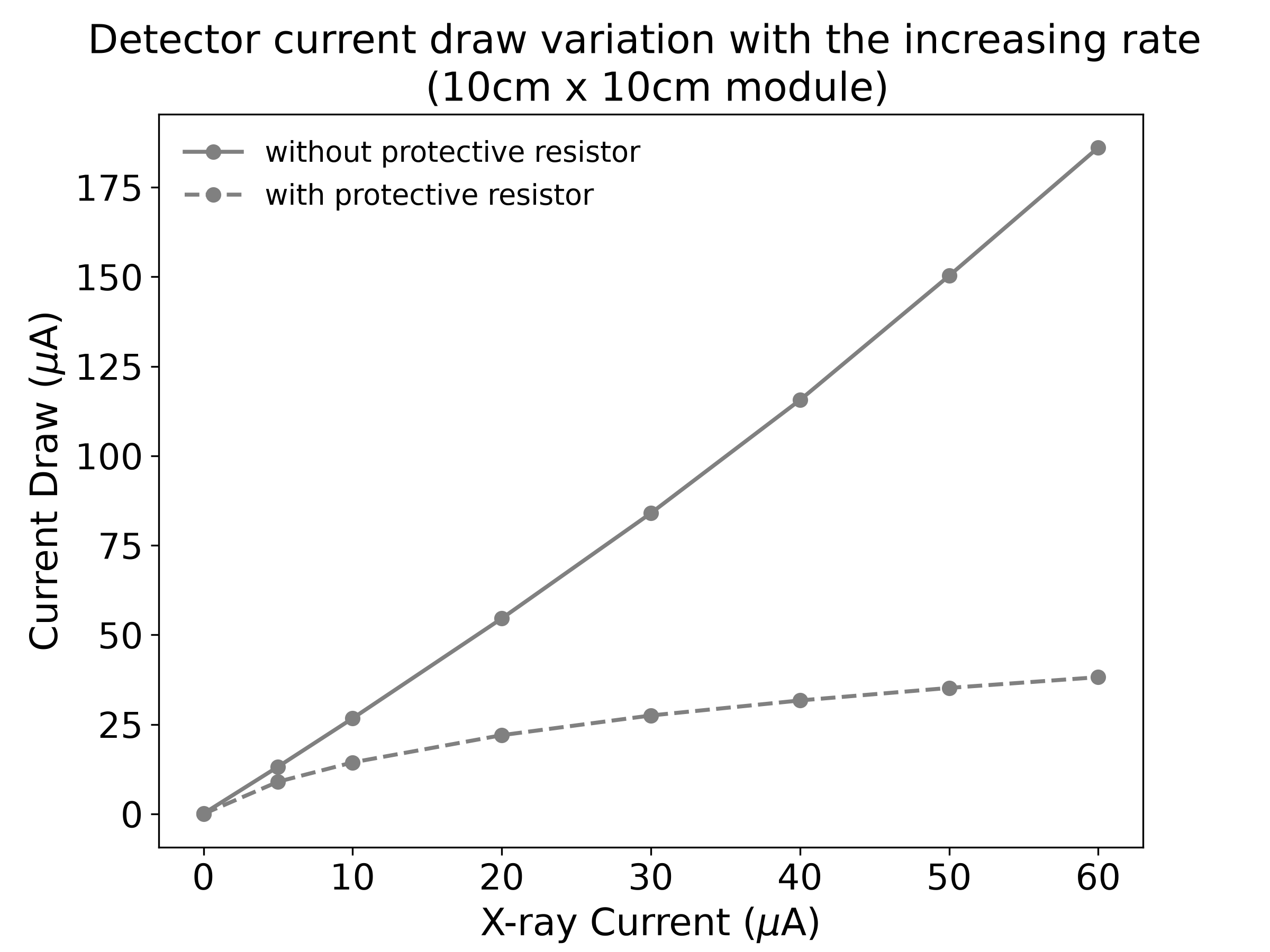}
    \caption{A comparison of the GEM detector current draw as a function of X-ray source current for two different protective current limiting resistor configurations with an individual channel power and a \qtyproduct{10x10}{\cm} GEM module. With large value protective resistors (1$~\mathrm{M\Omega}$) installed (dashed line), the current draw shows clear non-linearity and significantly lower compared to without (solid line).}
    \label{fig:Excess_current_10x10}
\end{figure}

To mitigate this rate-dependent gain loss, the high-voltage system was upgraded from the passive divider to an individual-channel configuration, allowing the electrode voltages to be controlled independently (\cref{fig:divider_and_individual_scheme}). X-ray studies demonstrated that this configuration produced a response much closer to the expected linear behavior. Further improvement was obtained by correcting the applied voltages for the voltage drops across the protective resistors on each channel (\cref{fig:Excess_current_50x60,fig:Excess_current_10x10}).

Several high-voltage supply types were used during the SBS experimental program after implementing the individual channel high-voltage scheme, including CAEN HiVolta DT1415E modules and CAEN A1515BTG and A1515BTGHP~\cite{CAEN_A1515B} modules housed in SY5527 crates~\cite{CAEN_SY5527}. The A1515BTGHP modules were preferred for high-rate operation because they offer higher per-channel current and power limits. The GEM high voltages were remotely controlled and monitored through the EPICS slow-control system~\cite{EPICSdocs}. Operating voltages for each detector were determined through efficiency plateau studies described in \cref{subscetion:Plateau_Studies}. 

\subsection{Low-Voltage Power Supply System}

As described in \cref{sec:Readout scheme}, the front-end electronics for the GEM detector were APV25 cards. These APV25 cards were organized in various quantities on backplanes. Successful readout of the GEM modules requires low-voltage power supplies to energize both the APV25 front-end cards and their distribution backplanes. Each APV25 consumes approximately \(296~\mathrm{mW}\) under nominal conditions~\cite{APVmanual}. The required $\SI{1.25}{V}$, $\SI{2.5}{V}$, and ground levels were generated locally on the GEM layers using radiation-hard LHC4913 regulator chips~\cite{LV_regulator_technote}, which lowered a DC input of \(\SI{5}{V}\). Early operation used custom Jefferson Lab low-voltage supplies to provide the input voltage, but periodic AC-to-DC converter failures motivated the transition to TDK-Lambda GEN 10-330-LAN3P480 supplies~\cite{TDKLambda}. These were installed in locations with lower radiation exposure in the experimental hall, to improve reliability.

\subsection{Gas Distribution System}

Two custom gas mixing and distribution systems were built for the GEM modules installed in both the BigBite and Super Bigbite Spectrometers. The key components and operating principles of both gas systems are largely identical. 

\begin{figure}[ht!]
    \centering
    \includegraphics[width=0.95\linewidth]{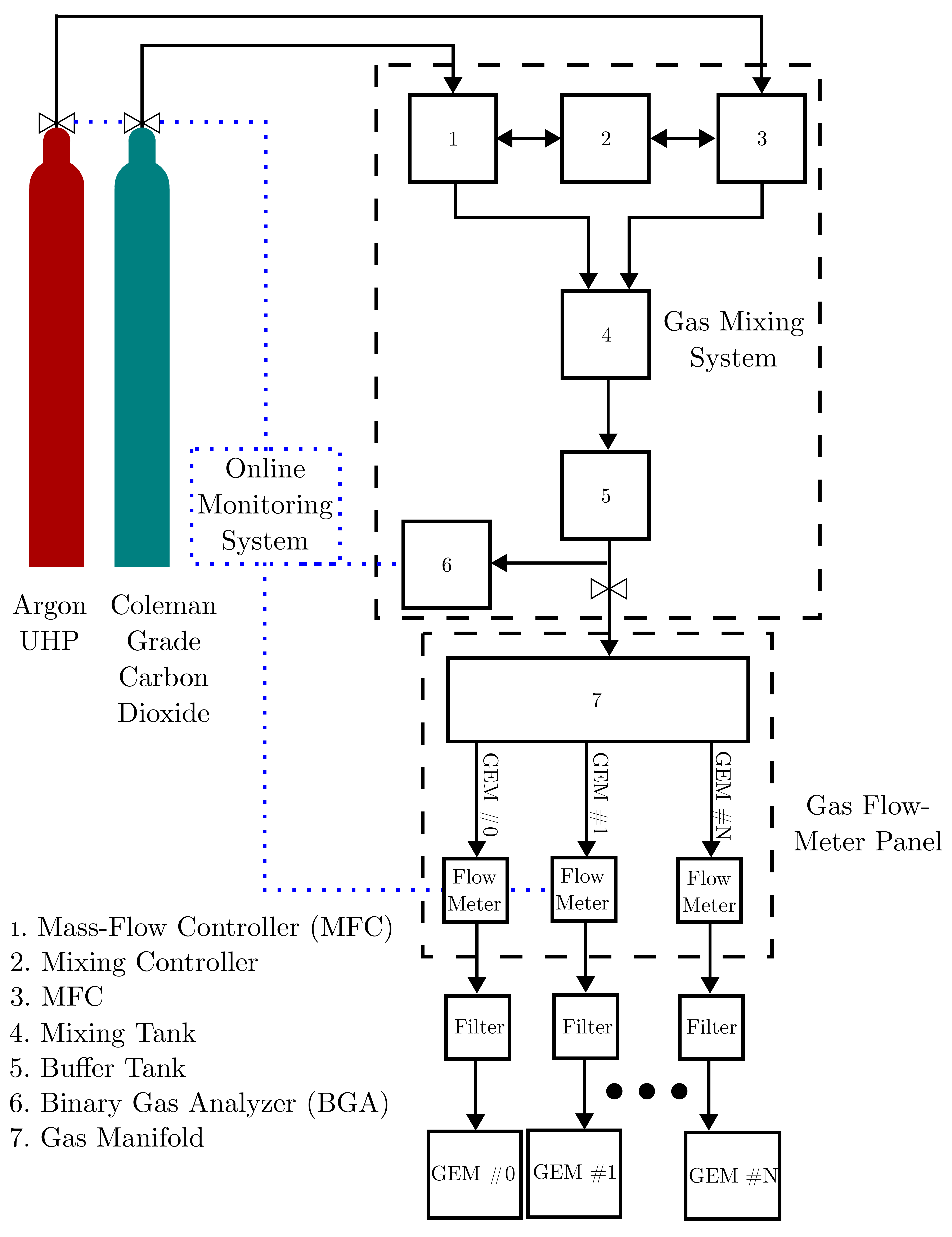}
    \caption{ A schematic of the key components of a gas distribution system during the SBS experimental program. Adapted from Ref.~\cite{EWertz}.}
    \label{fig:GEMGas}
\end{figure}

During in-beam operations and cosmic-ray testing, a typical gas mixture of Ar/CO$_2$ with the molar ratio of 75/25 was used for the GEM detectors of the SBS program. A schematic of the key components of a Hall A GEM gas system is presented in \cref{fig:GEMGas}.

Most components were located outside the beam enclosure, allowing for routine access and minimizing radiation exposure. The gas flow meter panel that provided gas lines to each GEM module was located within the experimental hall. Separate cylinders of ultra-high-purity argon and Coleman-grade carbon dioxide were connected to individual Mass Flow Controllers (MFCs), which regulated the relative flow rate into the mixing tank. The MFCs were controlled by the Mixing Controller module. The MFCs and the mixing controller ensure that the desired gas molar ratio of Ar/CO$_2$ is produced, within less than a 1\% uncertainty. The gas mixture ratio was constantly monitored (to an uncertainty within 0.01\%) by a Stanford Research Systems Binary Gas Analyzer (BGA) Model BGA244 \cite{BGA244m} connected to the outlet of the buffer tank. Throughout the SBS experiments, the BGAs were calibrated using multiple pre-mixed Ar/CO$_2$ gas cylinders supplied in different batches from the same distributor. A large-volume buffer tank, downstream of the mixing tank, holds the produced gas mixture and reduces fluctuation in pressure and mixture ratio. The buffer tank connects directly to the gas manifold panel inside the experimental hall, which contains all of the gas flow meters. Each individual GEM module (gas volume) is connected to its own flow meter. Gas filters with a filtration of 0.2-micron or finer (Mattheson 6164-P4FF and Parker 02FA06A) were installed downstream of each flow meter, and any remaining particulates from the gas mixture were filtered before entering the GEM modules. 

All major components, including pressure in each cylinder, BGA read back, and gas flow, were monitored via EPICS. The high-voltage system was programmed to turn off if the gas mixture ratio or gas flow rate deviated from safe operating conditions.

\section{RF shielding} \label{sec:GEM shielding} During the initial cosmic-ray commissioning of the SBS GEM modules, significant CM fluctuations of the raw signals were observed while the APV25 cards were connected to the readout strips. CM fluctuations were observed to be between 60--80 ADC units on average, with certain extreme instances where the fluctuation was over 100 ADC. A signal caused by a minimum ionizing particle in
a GEM module is expected to have a value no larger than a couple hundred ADC units and the
typical pedestal rms values have a mean of 10 ADC units. In this context, CM
fluctuations with the values as described would have a substantial effect on the noise profile of a given GEM module. In the online CM algorithm as described in \cref{sec:Readout scheme}, a sufficiently broad CM distribution could cause strips with signals in the APV25 to bias the calculation towards more positive values. 

\begin{figure}[htbp!]
    \centering
    \includegraphics[width=1.0\linewidth]{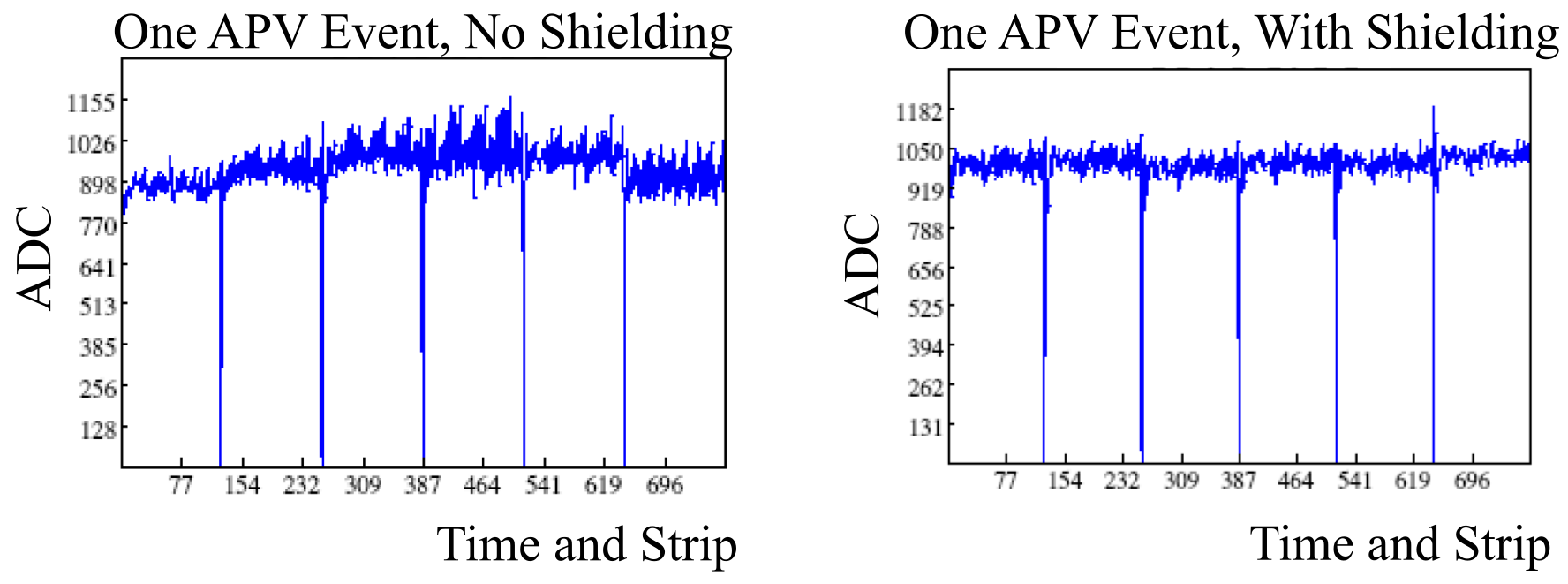}
    \caption{ An example of the effect of shielding on a single event GEM raw data frame. The key improvement is that the variation in the CM is significantly less after RF shielding. The ADC value is presented on the vertical axis and the strip number over the 6 time samples for the event is presented on the horizontal axis. Adapted from Ref.~\cite{EWertz}.}
    \label{fig:CM_fluctuation_single_event}
\end{figure}

Potential sources for these CM fluctuations were investigated, including power supplies connected to GEM readout electronics, ground loops, and interference from radio frequency (RF) electromagnetic waves in the surrounding environment. It was inferred that the most significant source for these excessive CM fluctuations was RF interference. In order to eliminate the induced electric field within the GEM modules by RF waves, the GEM modules were instrumented with \emph{Faraday Cage} (RF shielding) structures.

\begin{figure}[htbp!]
    \centering
    \includegraphics[width=0.95\linewidth]{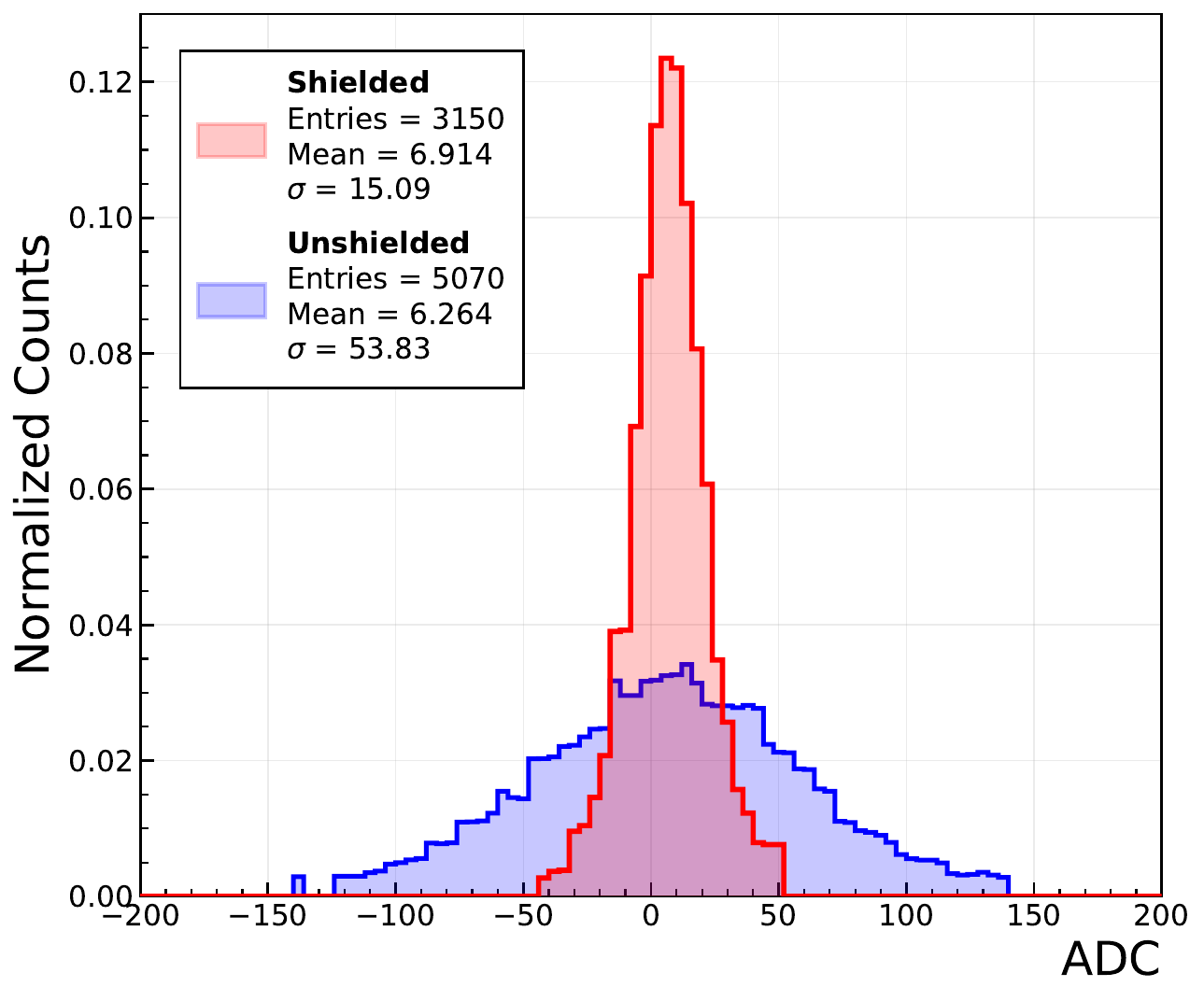}
    \caption{An example of a typical CM distribution comparison for a single APV25 card with (orange) and without RF shielding (blue).}
    \label{fig:CM_fluctuation_spectra_compare}
\end{figure}

\Cref{fig:CM_fluctuation_single_event} shows the effect of shielding on the fluctuation of the CM on a single APV25 card across all six time samples for a single event. It is evident from \cref{fig:CM_fluctuation_single_event} that the variation of the CM is more uniform after the shielding is installed. \Cref{fig:CM_fluctuation_spectra_compare} shows the effect of shielding on a single APV card, considering several thousand events. These histograms are generated by subtracting the calculated CM of the first time sample of a given event from the calculated CM of the next five time samples. This observable characterizes the variation of the CM within the six-sample readout window. The key factor for comparison is the standard deviation of these distributions. By comparing these CM distributions and the corresponding standard deviations for many APV25 cards before and after shielding, CM fluctuations were observed to have been reduced by a factor of 3--4.

\begin{figure}[htbp!]
    \centering
    \includegraphics[width=0.5\linewidth]{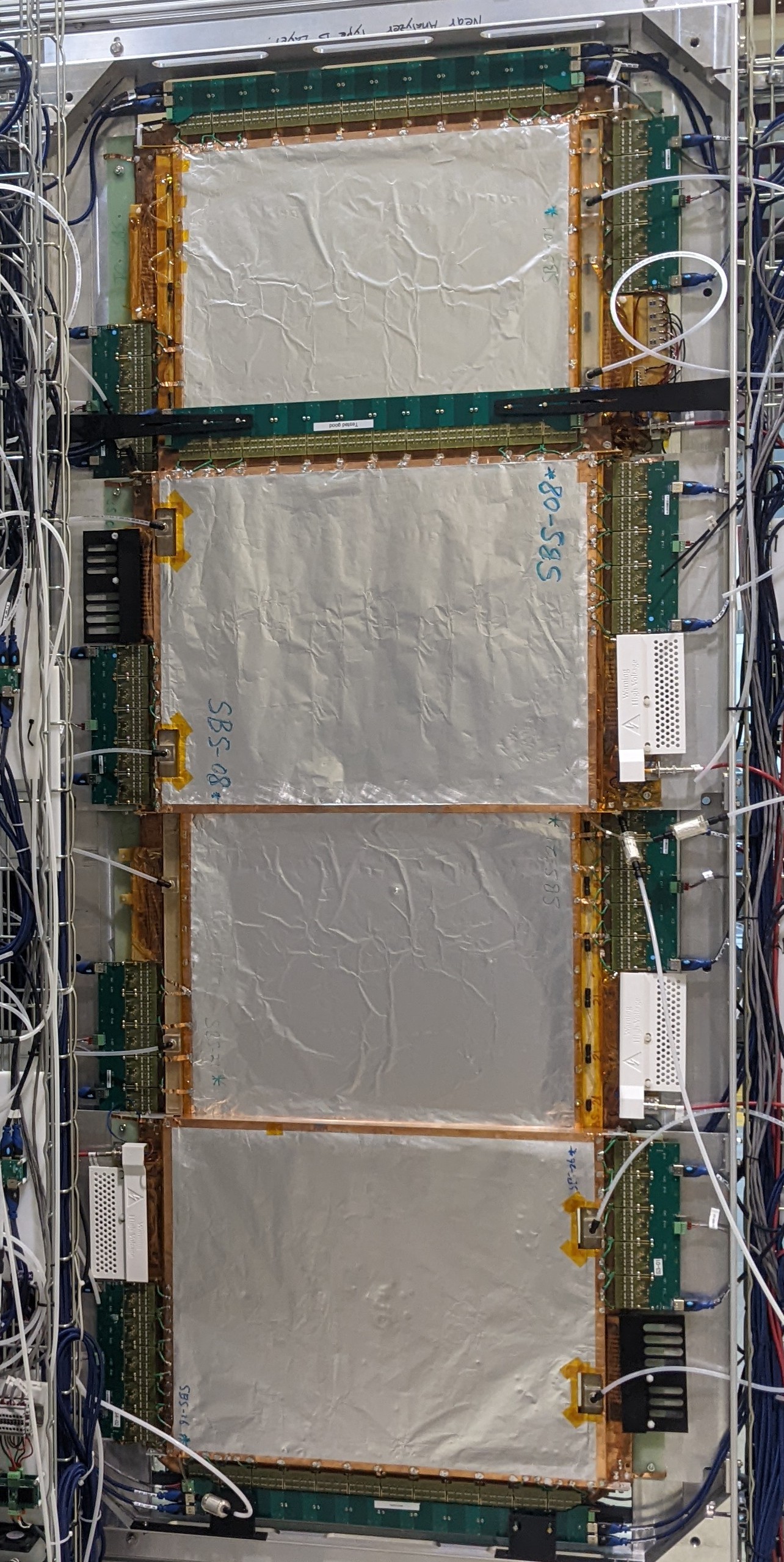}
    \caption{One XY GEM layer with RF shielding installed on every module.}
    \label{fig:XY_layer_with_shielding}
\end{figure}

\Cref{fig:XY_layer_with_shielding} shows an XY GEM layer equipped with shielding. Each different type of SBS GEM module has unique specifications for the final mechanical design of the electromagnetic shielding. However, components and features which are common for all shielding equipped to the SBS GEM layers, for production conditions, will be described. 

The support structure of the shielding was made of U channel aluminum bars; four bars were machined to be fastened together in a rectangular shape to mimic the perimeter of the GEM module frames. The primary component of the electromagnetic shield was a \SI{50}{\um} thick aluminum foil stretched across and connected to the aluminum support using conductive copper tape. The foil shields cover the GEM active area and introduces extraneous material to the tracking system, exacerbating multiple scattering effects and degrading the overall angular resolution. The thickness and material type were therefore chosen so that RF signals over a wide range would be shielded while minimizing the material budget. Aluminum foil with a thickness of \SI{50}{\micro\metre} -- matching the skin depth for \SI{3}{\mega\Hz} EM waves -- was selected as it provides sufficient shielding for frequencies above a few megahertz. Compared to alternatives such as copper, aluminum contributes to a significantly lower material budget, reducing the radiation length by more than a factor of six due to its lower atomic number and density. Consequently, the addition of two \SI{50}{\micro\meter} aluminum foils increases the baseline radiation length of the GEM module (approximately 0.5\%) by a moderate yet acceptable 0.1\%, to a total of 0.6\%.

Depending on the type of GEM module, a 1--3 cm gap is left between the aluminum foil and the gas window of the GEM module to suppress electrostatic interactions. Kapton tape and Kapton film, 3 mil and 5 mil thick, respectively, were included as necessary for electrical isolation between the shielding and the GEM module. Two shield frames were installed on either side of each GEM module and electrically connected using standard 20 AWG electrical wire soldered to the copper tape surrounding the frames. The number of wires connecting the two shield frames was at least equal to the total number of APV25 cards connected to the corresponding GEM module. Similar electrical wires were soldered to the copper tape and connected to the ground of the APV25 cards by mechanical fasteners.

\section{Cosmic-Ray Testing} \label{sec:Cosmic testing} 
\begin{figure}[t!]
    \centering
    \includegraphics[width= 0.8 \linewidth]{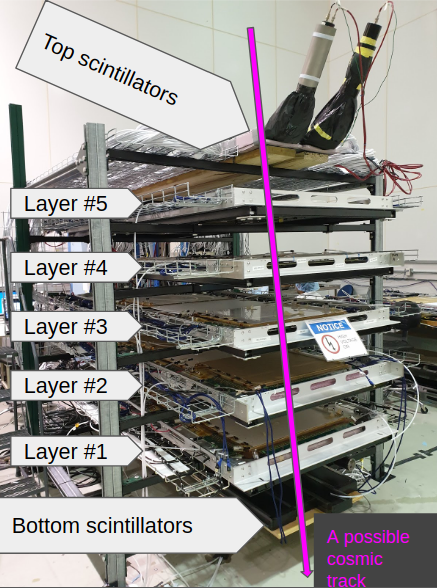}
    \caption{Test stand constructed for cosmic-ray data collection and track-based detector analysis. In this picture 5 XY GEM layers are present, as well as the trigger scintillators at the top and bottom of the test stand.}
    \label{fig:cosmic_test_stand}
\end{figure}

Extensive cosmic-ray testing was performed to characterize the SBS GEM detectors prior to installation in the experimental hall. The active area and track-based efficiency were evaluated for each GEM module. A mechanical test stand (\cref{fig:cosmic_test_stand}) was constructed to hold up to five GEM layers for the purpose of collecting data from cosmic-rays, allowing for the evaluation of key detector performance parameters. Each GEM layer was installed horizontally and sequential GEM layers were stacked vertically on the stand. Two layers of scintillators were installed, one above and one below the GEM stack, to trigger on vertical cosmic ray trajectories traversing the test stand.

\subsection{Track Reconstruction}
The track reconstruction algorithm used for cosmic-ray commissioning data was based on a standalone framework~\cite{SBSGEM_standalone} developed for low-luminosity conditions. Reconstruction consists of three major stages: one-dimensional cluster finding, two-dimensional hit formation, and track reconstruction. One-dimensional cluster finding and two-dimensional hit formation are performed on a per-module basis, whereas track reconstruction incorporates hits from the full set of tracking layers.

\subsubsection{1D Cluster Finding}\label{sec:1D_cluster_finding}
The signals recorded on the two strip projections were processed independently to form one-dimensional clusters. The two projections are denoted \textit{U} and \textit{V}; these labels refer to the two nonparallel position coordinates measured by the strip planes and need not coincide with the global Cartesian \textit{X} and \textit{Y} axes. The angle between the two strip projections was configurable to accommodate various readout types used in the SBS GEM modules.

For each strip (\textit{i}), the charge ($q_i$) was obtained by summing the six digitized APV25 samples,
\begin{equation}
    q_i=\sum_{s = 0, \dots, 5}A_{i,s},
\end{equation}
where $A_{i,s}$ is the ADC value of the \textit{s}'th sample. Before clustering, strips were required to satisfy minimum thresholds on both the largest individual ADC sample and the sum of all samples. The position of the maximum sample was also required to lie within an accepted sampling window, as determined by the trigger-latency.  These requirements suppressed strips dominated by electronic noise or signals uncorrelated with the trigger signal. 

An amplitude-weighted time was calculated for each retained strip according to:
\begin{equation}\label{eq:strip_time}
    t_i = \frac{\sum_{s=0,\dots,5} A_{i,s}t_s}{\sum_{s=0,\dots,5} A_{i,s}},
\end{equation}
where $t_s$ is the center time of the sample \textit{s}.

One-dimensional clusters were identified separately in the U and V projections. A retained strip was initially classified as a \emph{local maximum} when its integrated charge was greater than or equal to that of each of the two immediately adjacent strips. Each such local maximum was considered a potential seed for a one-dimensional cluster.

Though not frequent in cosmic data, a single contiguous group of fired strips can contain more than one local maximum, particularly in high-occupancy events when signals from nearby avalanches overlap. To distinguish significant secondary peaks from fluctuations on the shoulder of a larger peak, the prominence of each secondary maximum was evaluated. For a secondary peak of charge $q_{\rm peak}$, the prominence was defined as
\begin{equation}
    P = q_{\rm peak} - q_{\rm valley},
\end{equation}
where $q_{\rm valley}$ is the minimum charge between the candidate secondary peak and a neighboring higher peak. A secondary maximum was suppressed when its prominence was insufficient relative to the expected electronic noise or when the prominence was less than a specified fraction of the peak amplitude. The prominence test was applied only to peaks sufficiently close to a higher maximum to produce overlapping cluster windows.

\begin{figure}[t!]
    \centering
    \includegraphics[width= 0.95 \linewidth]{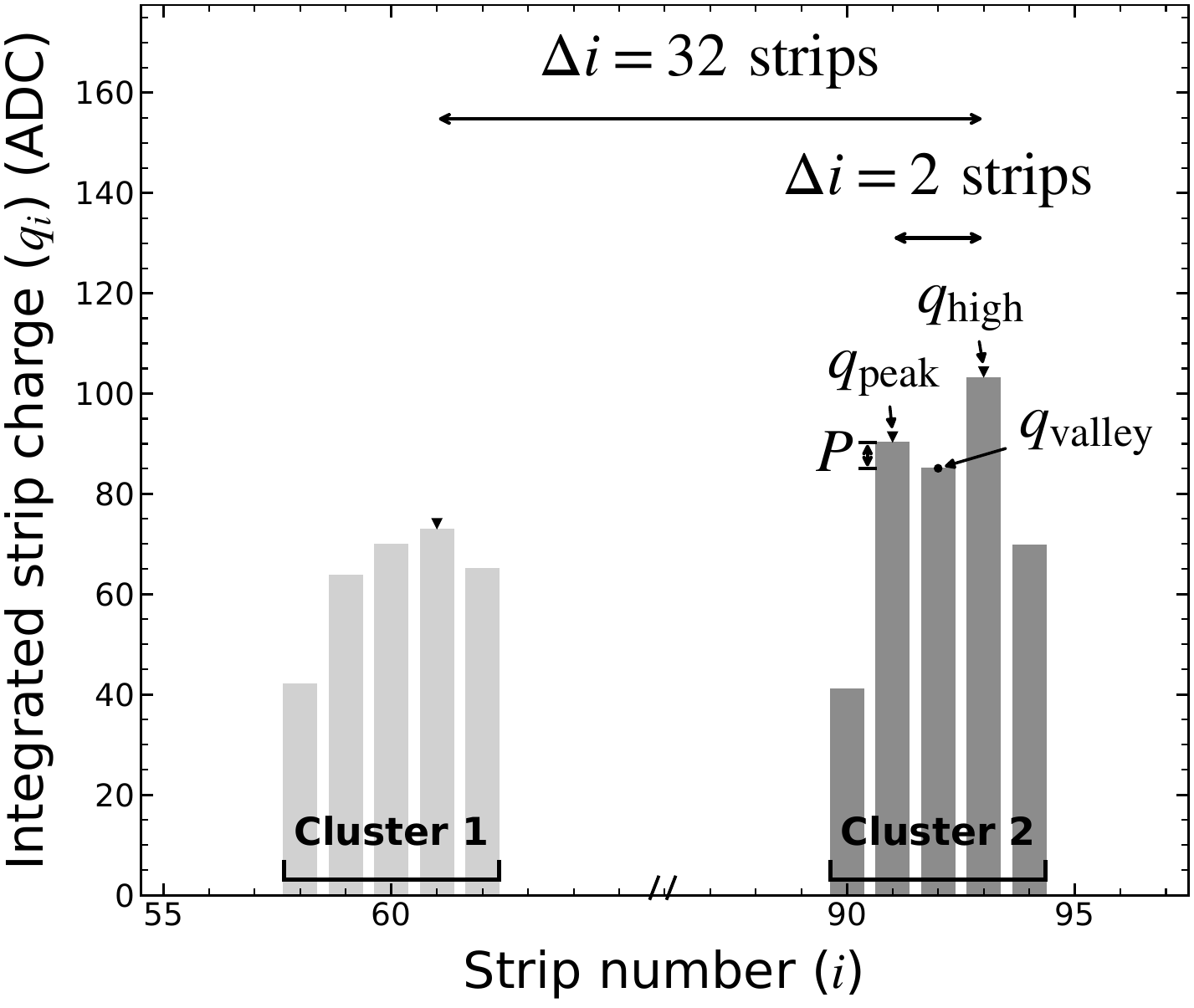}
    \caption{An example cosmic event demonstrating the one-dimensional clustering}
    \label{fig:cluster_splitting}
\end{figure}

\Cref{fig:cluster_splitting} shows a representative cosmic-ray event illustrating the one-dimensional GEM cluster finding procedure. The isolated maximum on the left is separated from the dominant maximum on the right by $\Delta i = 32$ strips and is therefore reconstructed as a separate cluster. Within the right-hand cluster, two local maxima are separated by only $\Delta i = 2$ strips. Because the secondary maximum has only a small prominence, $P = q_{\mathrm{peak}} - q_{\mathrm{valley}}$, it is not retained as an independent cluster and is instead merged with the neighboring higher maximum. This event therefore illustrates how both peak separation and prominence are used to distinguish physically separate clusters from nearby fluctuations within a single cluster.

For each accepted maximum, neighboring retained strips were collected in both directions until a missing strip was encountered or the maximum configured distance from the seed was reached. When the windows surrounding two accepted maxima overlapped, the charge on a shared strip was divided between the corresponding clusters. For a cluster seeded by maximum \textit{m}, the relative contribution of strip \textit{i} was estimated using the weight:
\begin{equation}
    w_{mi} = \frac{q_m}{1+ \left[ (u_i-u_m)/\sigma_{\rm shape} \right]^2 },
\end{equation}
where $q_m$ is the charge of the seed strip, $(u_i-u_m)$ is the separation between the strip and the seed, and $\sigma_{\rm shape}$ is a characteristic transverse signal width. The fraction of the strip charge assigned to cluster $m$ is given by:
\begin{equation}
    f_{mi} = \frac{w_{mi}}{\sum_j w_{ji}},
\end{equation}
where the sum includes all nearby cluster seeds (\textit{j}) that contribute to strip \textit{i}. The charge assigned to the cluster \textit{m} from strip \textit{i} was consequently $f_{mi}q_i$. The same fraction was applied to every ADC time sample, preserving the pulse shape associated with each reconstructed cluster. This charge-sharing procedure allows multiple spatially overlapping clusters to be reconstructed without assigning the full charge of a shared strip to every peak.                  

The mean position of each one-dimensional cluster was obtained from a charge-weighted centroid,
\begin{equation}
    u_{\rm cl} = \frac{\sum_i u_i q_i^{\rm cl}}{\sum_i q_i^{\rm cl}},
\end{equation}
where $q_i^{\rm cl} = f_{mi}q_i$ is the portion of the strip \textit{i}'s charge assigned to the cluster. The strip-center coordinate ($u_i$) is calculated from the strip number, number of strips in the projection, and the strip pitch.

In addition, other important quantities such as cluster charge ($Q_m$), and cluster time ($t_m$) are defined as:
\begin{equation}\label{eq:Cluster_charge}
    Q_m = \sum_i {q_i^{\rm cl}},
\end{equation}
\begin{equation}\label{eq:Cluster_time}
    t_m = \frac{\sum_i{q_i^{\rm cl}t_i}}{Q_m},
\end{equation}
where $t_i$ is defined in the above \cref{eq:strip_time}.

\subsubsection{2D Hit Construction}\label{sec:2Dhit_formation}
After clustering the two projections independently, two-dimensional hit candidates were constructed within each GEM module. Every \textit{U}-projection cluster was initially paired with every \textit{V}-projection cluster. The APV25 pulse shapes provided additional information for identifying the physically correct $(U,V)$ cluster association. For each candidate, a Pearson correlation coefficient was calculated between the cluster summed ADC samples in the two projections:

\begin{equation}
    \rho_{\rm cl} = \frac{\sum_{s=0,\dots,5} (U_s - ~\overline{U})(V_s - ~\overline{V})}{\sqrt{\left[\Sigma_{s=0,\dots,5}(U_s-\overline{U})^2\right]\left[\Sigma_{s=0,\dots,5}(V_s-\overline{V})^2\right]}},
\end{equation}
where $U_s$ and $V_s$ are the cluster-summed signals in sample \textit{s}. A second correlation coefficient, $\rho_{\rm max}$, was calculated from the sample sequences of the two maximum-charge strips in the two clusters. True $(U,V)$ associations are expected to exhibit similar pulse shapes and thus have Pearson correlation coefficients closer to unity, because the signals on the two projections originate from the same GEM avalanche.

The charge and time compatibility of each candidate were characterized using 

\begin{equation}
    Q_{\rm hit} = \frac{Q_U+Q_V}{2},
    \qquad
    A_Q = \frac{Q_U-Q_V}{Q_U+Q_V},
\end{equation}
and

\begin{equation}\label{eq:Hit_time_timediff}
    t_{\rm hit} = \frac{t_U+t_V}{2},
    \qquad
    \Delta t = t_U-t_V,
\end{equation}
where $Q_U$, $Q_V$, $t_U$, and $t_V$ are the charges and reconstructed times of the component clusters.

Ambiguous $(U,V)$ combinations were reduced through a sequential pruning procedure. The quantities considered, in order, were the geometric-mean charge ($\sqrt{Q_UQ_V}$), the time difference ($\Delta t$), the charge asymmetry ($|A_Q|$), the maximum-strip correlation coefficient ($\rho_{\rm max}$), and the cluster-summed correlation coefficient ($\rho_{\rm cl}$). At each stage, candidates failing the applicable criterion were removed if at least one of the currently surviving candidates in the module passed the criterion. If no candidates passed a particular criterion, that criterion was not used to reject any candidates in the module. This fallback procedure preserved at least one possible hit association for subsequent resolution by the track-finding algorithm, and prevented candidate rejection from a poorly configured criterion. The numerical thresholds were configurable and were selected for the detector operating conditions and event sample.

The local Cartesian position of each candidate was obtained by combining its measured \textit{U} and \textit{V} coordinates. For general non-orthogonal readout coordinates,

\begin{equation}
    U=P_{xu}X+P_{yu}Y,
    \qquad
    V=P_{xv}X+P_{yv}Y,
\end{equation}
where the coefficients \textit{P} are determined by the configured orientations of the two strip projections. Inverting these relations gives

\begin{equation}
    X = \frac{P_{yv}U-P_{yu}V}{P_{xu}P_{yv}-P_{yu}P_{xv}},
    \qquad
    Y = \frac{P_{xu}V-P_{xv}U}{P_{xu}P_{yv}-P_{yu}P_{xv}}.
\end{equation}

Finally, the reconstructed hit position in the local detector coordinate system was transformed to a global tracker coordinate system by taking into account module alignment, rotation, and translation with respect to a single reference GEM module within the tracking system. The offset parameters, with respect to the chosen reference GEM module, necessary to perform the module to global coordinate transformation were determined by a software alignment procedure \cite{SBSGEM_standalone} using large samples of clean straight-line cosmic ray trajectories.

\subsubsection{Track Formation}\label{sec:Track_formation}
Track formation is performed by combining clusters across multiple GEM layers. Only events with hits in at least three distinct layers are considered. Candidate tracks are formed by considering all possible combinations of one 2D hit candidate per GEM layer potentially consistent with a straight-line track in 3D space. This ``brute force'' approach was computationally manageable under the low-occupancy conditions characteristic of cosmic-ray tests. Most potential hit combinations are eliminated from consideration by dividing each layer's active area into a uniform two-dimensional rectangular grid and collecting a list of hits in each grid bin. Pairs consisting of one hit from each of the two outermost layers were used to define straight-line projections to the inner layers, which in turn were used to select an appropriate subset of grid bin(s) and hits to be considered in each inner layer for a given candidate hit pair from the outermost layers, eliminating many obviously invalid combinations from consideration. 

For each candidate hit combination, the straight-line track parameters $(x_{0},\frac{dx}{dz}, y_0, \frac{dy}{dz})$ were obtained from a least-squares fit. The quality of the fit is quantified using a $\chi^2$ per degree of freedom:
\begin{equation}
    \chi^2/\text{ndf} = \frac{1}{N_{\text{hits}}-2}\sum_{i=1}^{N_{\text{hits}}}{\left[ \frac{(x_{i}-x(z_{i}))^2}{\sigma_{x}^2} + \frac{(y_{i}-y(z_{i}))^2}{\sigma_{y}^2}\right]},
\end{equation}
where $(x_i,y_i,z_i)$ are the hit coordinates in 3D space, $\sigma_{x}$ and $\sigma_{y}$ represent estimated position resolutions, and 
\begin{equation}
    \begin{split}
        x(z_i) \equiv x_0 + \left(\frac{dx}{dz}\right)z_i,\\
        y(z_i) \equiv y_0 + \left(\frac{dy}{dz}\right)z_i,
    \end{split}
\end{equation}
are the projected positions of the straight-line track at each GEM layer.

On each track-finding iteration, the combination with the smallest $\chi^2/\text{ndf}$ is selected as the ``best'' candidate. If the best candidate's $\chi^2/\text{ndf}$ is below a user-defined maximum value, it is deemed a valid track, and its associated 2D hits and 1D clusters are removed from the candidate list. The procedure is then repeated until no additional valid tracks are found. 

On the first iteration of track-finding, hits are required on all GEM layers. To account for the possibility that one or more GEM layers are missing from otherwise valid tracks due to detection and/or reconstruction inefficiency and/or acceptance effects, the aforementioned track-finding procedure is repeated with decreasing minimum hit requirements until the user-specified minimum number of hits on a valid track is reached (3 in most of the tests described here). All possible combinations of layers are treated on an equal footing at any given minimum hit requirement, in the sense that the track candidate with the smallest $\chi^2/\text{ndf}$ is chosen as the best candidate, regardless of which layers are missing from the track.  

In the following sections, we present results from the XY GEM modules, which are representative of the performance of all SBS GEMs.

\subsection{X and Y Cluster Charge Sharing and ADC Asymmetry}
An important feature to identify true 2D hits among a large number of candidate 2D hits is the relationship between the ADC signal strength of the two 1D clusters in the two readout strip planes of a given 2D hit. The localized charge induced by the same avalanche cloud on the two overlapping readout strip planes should have a strong linear correlation. Such a correlation supports accurate 2D hit re-combination and false 2D hit rejection. This feature is even more important for high occupancy running conditions in SBS experiments to reduce the otherwise prohibitively large number of false 2D hit combinations that could be formed. 

\begin{figure}[!]
    \centering
    \includegraphics[width=\linewidth]{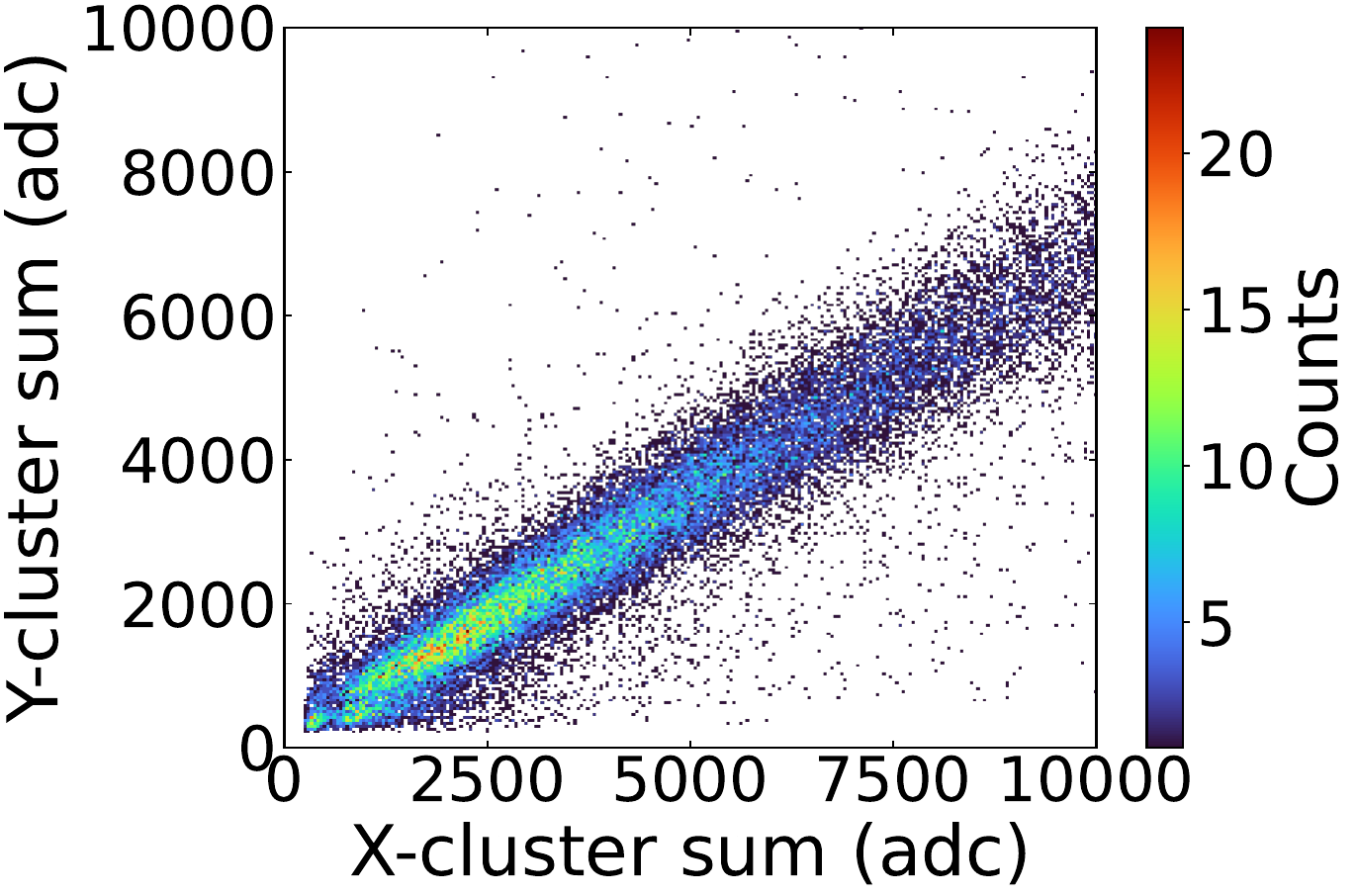}
    \caption{Y cluster sum ADC vs X cluster sum ADC.}
    \label{fig:YvsXclusterADC}
\end{figure}

\Cref{fig:YvsXclusterADC} shows the ADC correlation of the sum of the clusters for one of the XY GEM modules. Here, the $X(Y)$ cluster ADC sum is defined as the sum of all ADC values over all time samples from every strip in the cluster:
\begin{equation}
    Q_{X(Y)} = \sum_{i=1}^{N_{\text{strip}}}q^{X(Y)}_{i},
\end{equation}
where $q_{i}^{X(Y)}$ is the sum of the ADC units over all time samples of the strip \textit{i}, in $X(Y)$ readout plane, and the index \textit{i} runs over all strips in the cluster.

\begin{figure}[!]
    \centering
    \includegraphics[width=\linewidth]{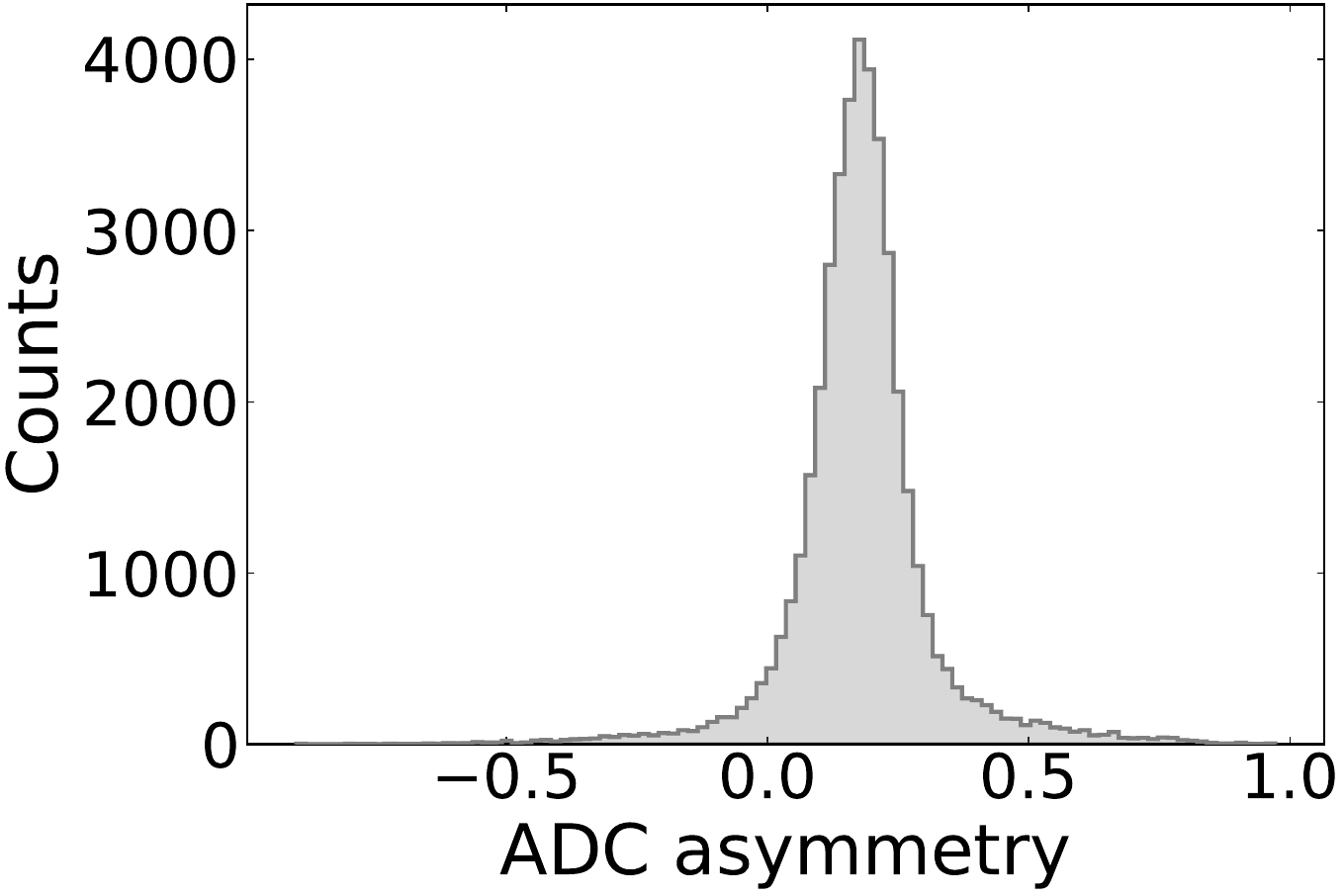}
    \caption{X and Y cluster sum ADC asymmetry.}
    \label{fig:ClusterADCasym}
\end{figure}

For the same set of 1D cluster combinations, \cref{fig:ClusterADCasym} shows the ADC asymmetry, defined as the difference-over-sum ratio of the \textit{X} and \textit{Y} cluster sum ADCs: 
\begin{equation}
    \text{Cluster ADC Asymmetry} = \frac{Q_{X}-Q_Y}{Q_X+Q_Y}.
\end{equation}

The strong linear correlation of the \textit{X} vs \textit{Y} cluster sum ADC and the narrow ADC asymmetry distribution are characteristic of a high-quality readout board. Exact correlations can vary depending on the GEM module and should be characterized and considered in experiment data analysis. 

The systematically lower cluster charge of the Y cluster of a hit compared to the X cluster is expected. The Y readout strip plane is located underneath the X readout strip plane and hence a lower fraction of charge is induced on the Y strip plane due to the shielding from the X strip plane.
\begin{figure*}[htpb!]
    \centering
\begin{subfigure}{0.45\linewidth}
    \includegraphics[width=\linewidth]{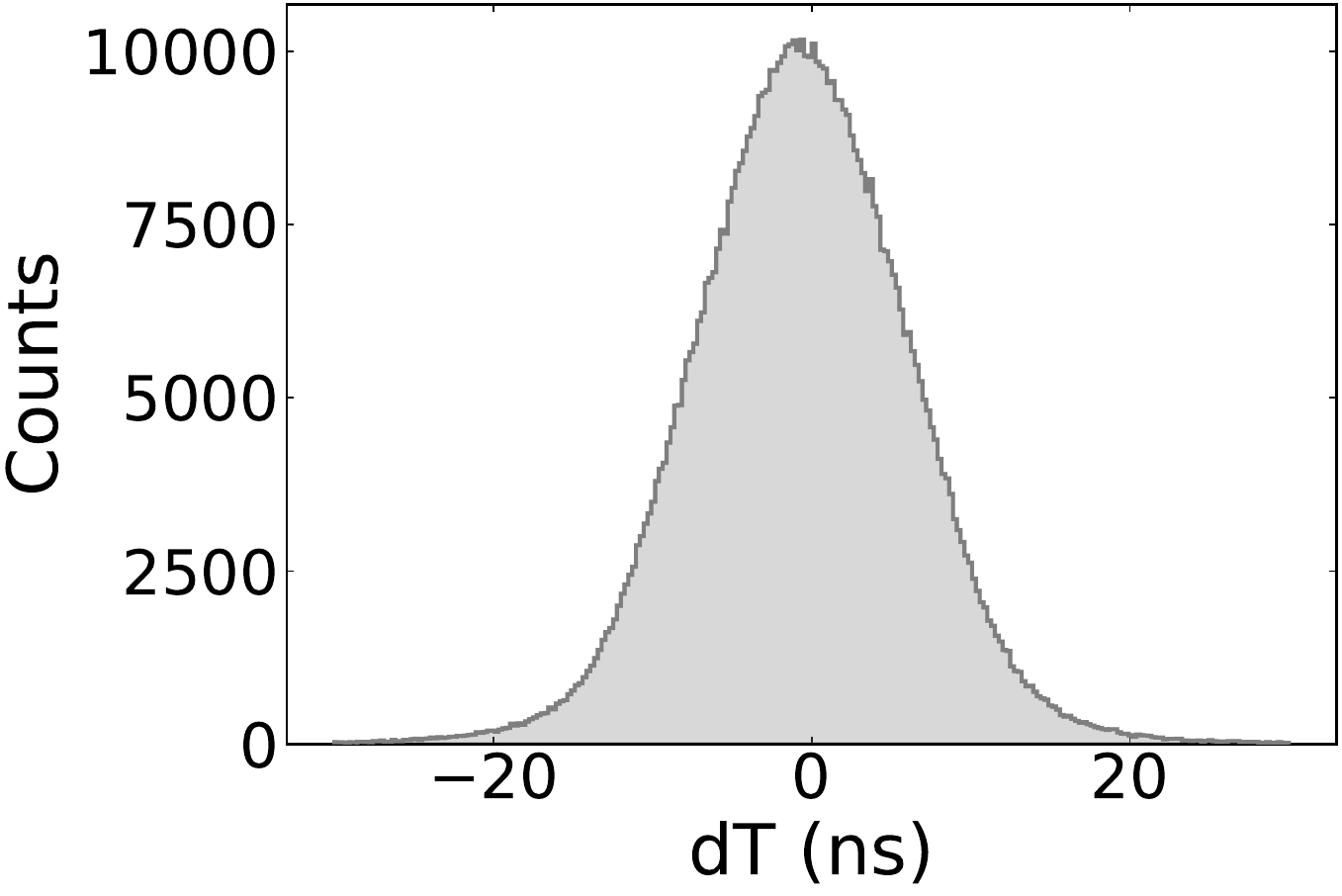}
    \caption{}
    \label{fig:XandYcluster_dT}
\end{subfigure}
\begin{subfigure}{0.45\linewidth}
    \includegraphics[width=\linewidth]{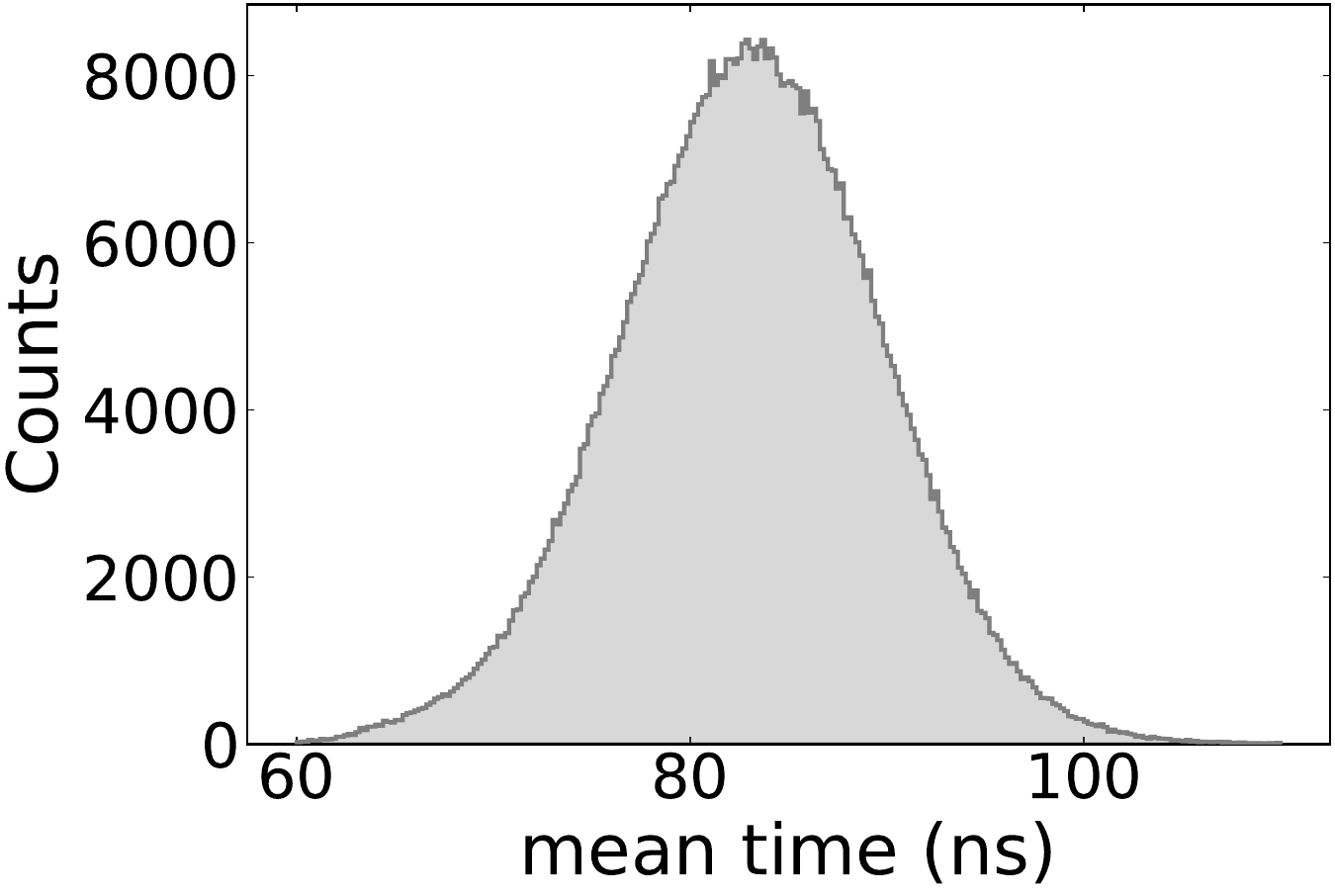}
    \caption{}
    \label{fig:XandYcluster_Tmean}
\end{subfigure}
\caption{\textit{X} and \textit{Y} cluster time difference (left), and the 2D hit time (right).}
\end{figure*}

\subsection{Timing Performance}
Although the nature of cosmic-ray data complicates an accurate time resolution analysis of GEMs, several existing metrics, such as the 1D cluster time (defined in \cref{eq:Cluster_time}) and 2D hit time (defined in \cref{eq:Hit_time_timediff}), can be used to gauge the timing performance of GEM modules. 

\Cref{fig:XandYcluster_dT} shows the difference between the \textit{X} and \textit{Y} 1D cluster times for 2D hits that form tracks in a single XY GEM module. During data analysis, hit combinations in the 2D hit formation procedure that fall outside a specified number of standard deviations from such distributions can be eliminated, as further described in \cref{sec:2Dhit_formation}. \Cref{fig:XandYcluster_Tmean} shows the distribution of the mean of \textit{X} and \textit{Y} cluster times, referred to as the 2D hit time, for this same data set. The sharpness of this 2D hit time distribution can serve as a proxy for the timing resolution of the GEM module.

\subsection{Spatial Resolution}

\begin{figure*}[t!] 
    \centering
    \includegraphics[width=0.9\linewidth]{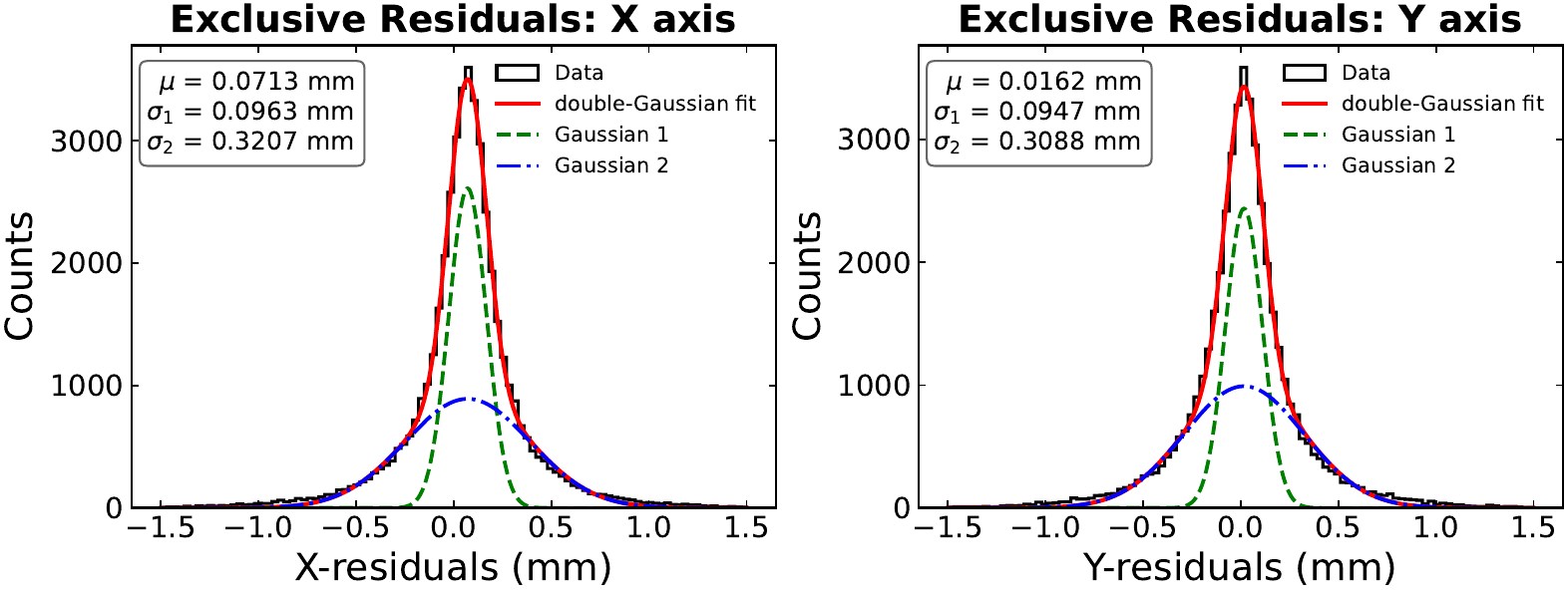}
    \caption{Exclusive residual distributions (black lines) along \textit{X} and \textit{Y} strip projections. The distributions are fitted with a double Gaussian function (red curves). The dashed-green curves are the narrower Gaussian function and the dashed-blue curves are the wider Gaussian function.}
    \label{fig:exclusive_resid_M8}
\end{figure*}

The spatial resolution of a GEM module is an important performance metric and is quantified using exclusive tracking residuals. An exclusive tracking residual is calculated by removing a 2D GEM hit, re-fitting the track with the remaining 2D hits, and taking the difference between the observed hit and the re-fitted track projection. Calculating this residual requires at least three additional 2D hits to remain on the track.\\
\Cref{fig:exclusive_resid_M8} shows the exclusive residual distributions along \textit{X} and \textit{Y} strip projections of an XY GEM module. Because these are cosmic ray measurements, the resulting resolution is subject to degradation compared to in-beam performance, due to the larger angular spread of the cosmic rays. In order to minimize this effect, cosmic rays that are within \SI{5}{\degree} of the normal direction to the GEM modules are used for this analysis. In addition, stringent track-quality criteria are applied, including a minimum of five 2D hits per track and a maximum $\chi^2/ndf$ of 25 for a fitted track, in order to minimize the inclusion of false-tracks. A double Gaussian function is fitted to the residual distributions to capture a more accurate fit for the sharper central peak and non-Gaussian tails. The standard deviations of the narrower Gaussian components along \textit{X} and \textit{Y} strip projections are \SI{96.3}{\um} and \SI{94.7}{\um}, respectively.

\begin{figure*}[tbp]
  \centering

  \begin{subfigure}[t]{0.49\textwidth}
    \centering
    \includegraphics[width=\linewidth]{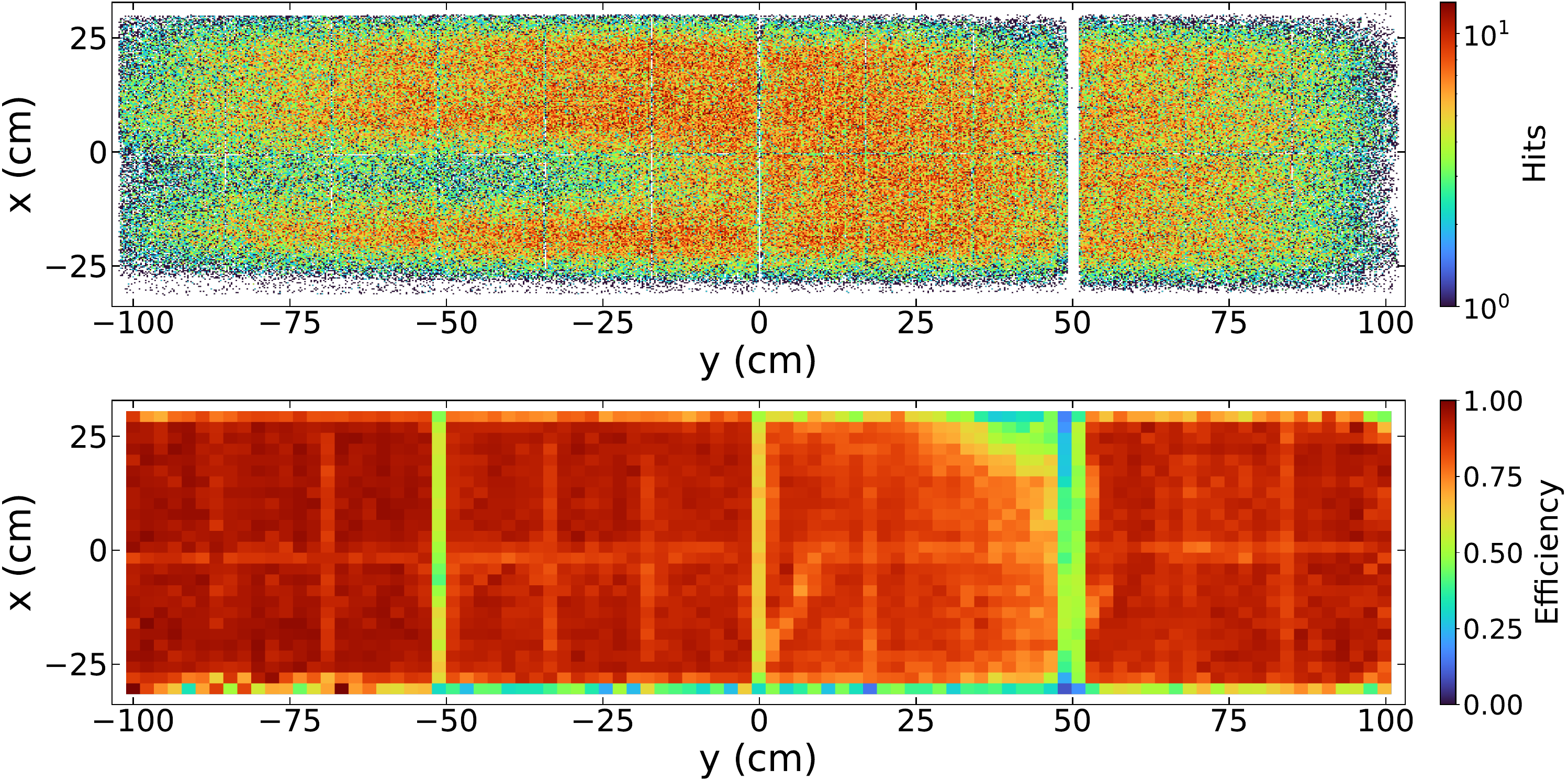}
    \caption{Layer 1}
    \label{fig:eff_L0}
  \end{subfigure}\hfill
  \begin{subfigure}[t]{0.49\textwidth}
    \centering
    \includegraphics[width=\linewidth]{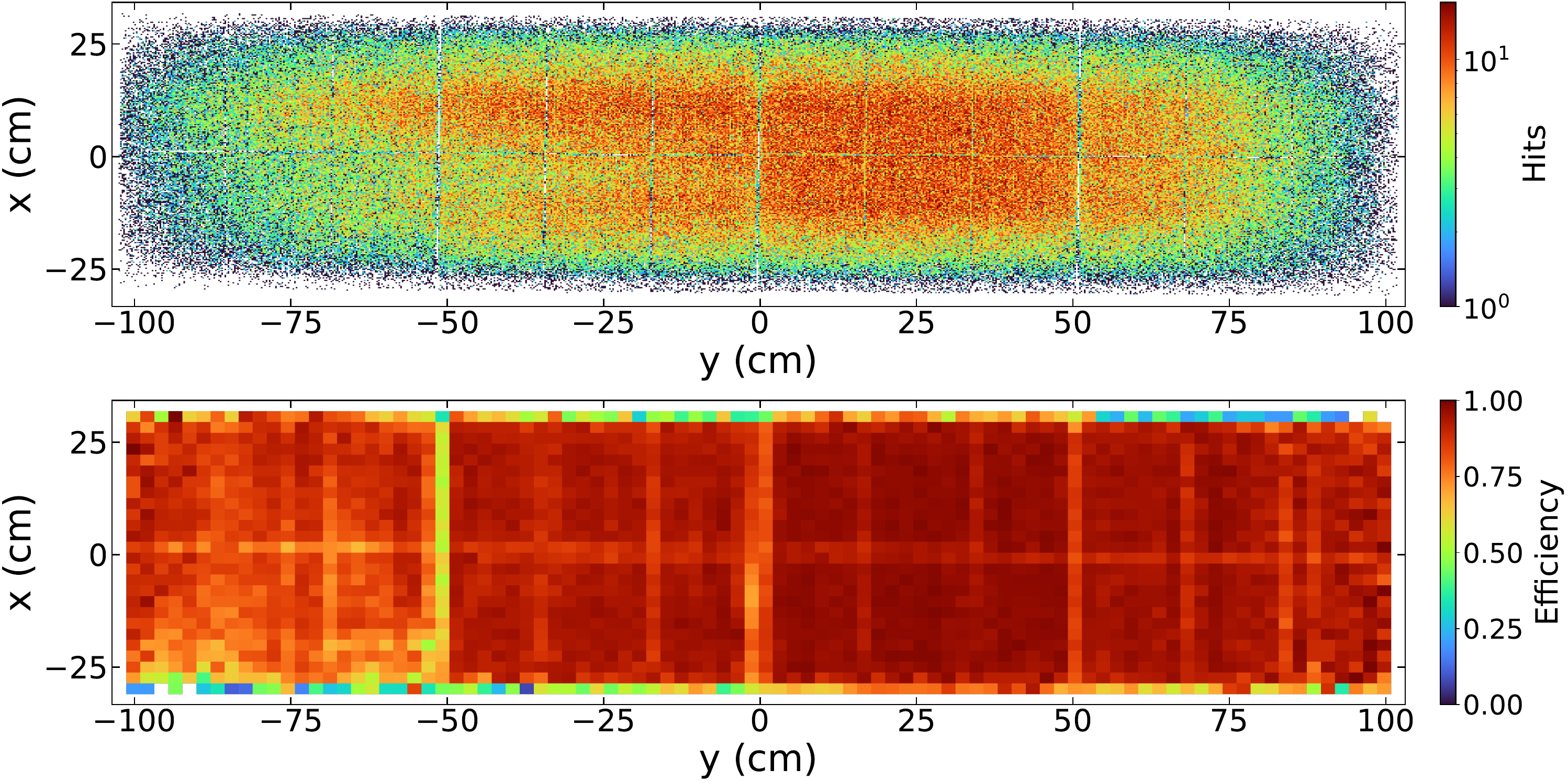}
    \caption{Layer 2}
    \label{fig:eff_L1}
  \end{subfigure}

  \vspace{1.0em}

  \begin{subfigure}[t]{0.49\textwidth}
    \centering
    \includegraphics[width=\linewidth]{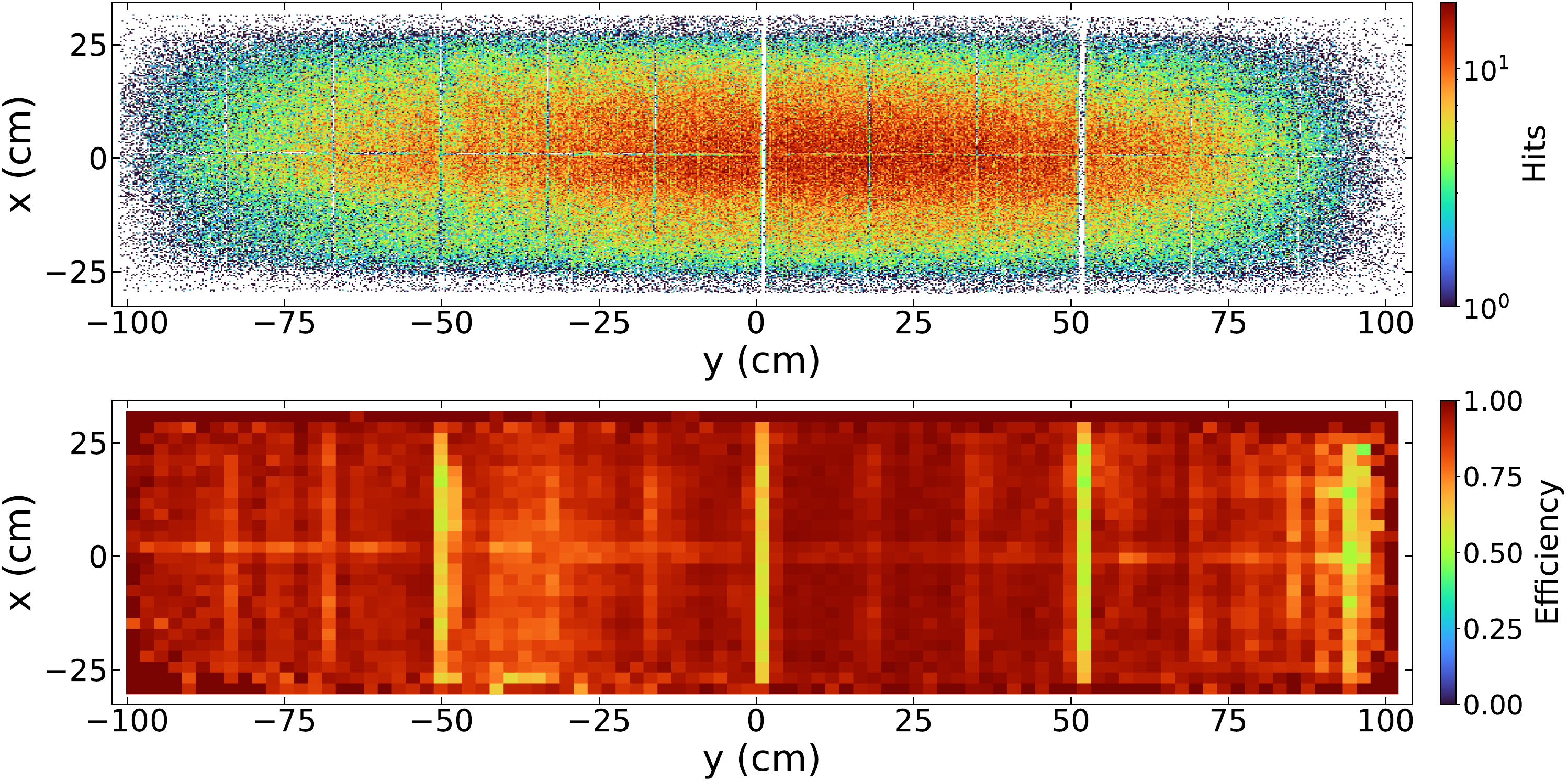}
    \caption{Layer 3}
    \label{fig:eff_L2}
  \end{subfigure}\hfill
  \begin{subfigure}[t]{0.49\textwidth}
    \centering
    \includegraphics[width=\linewidth]{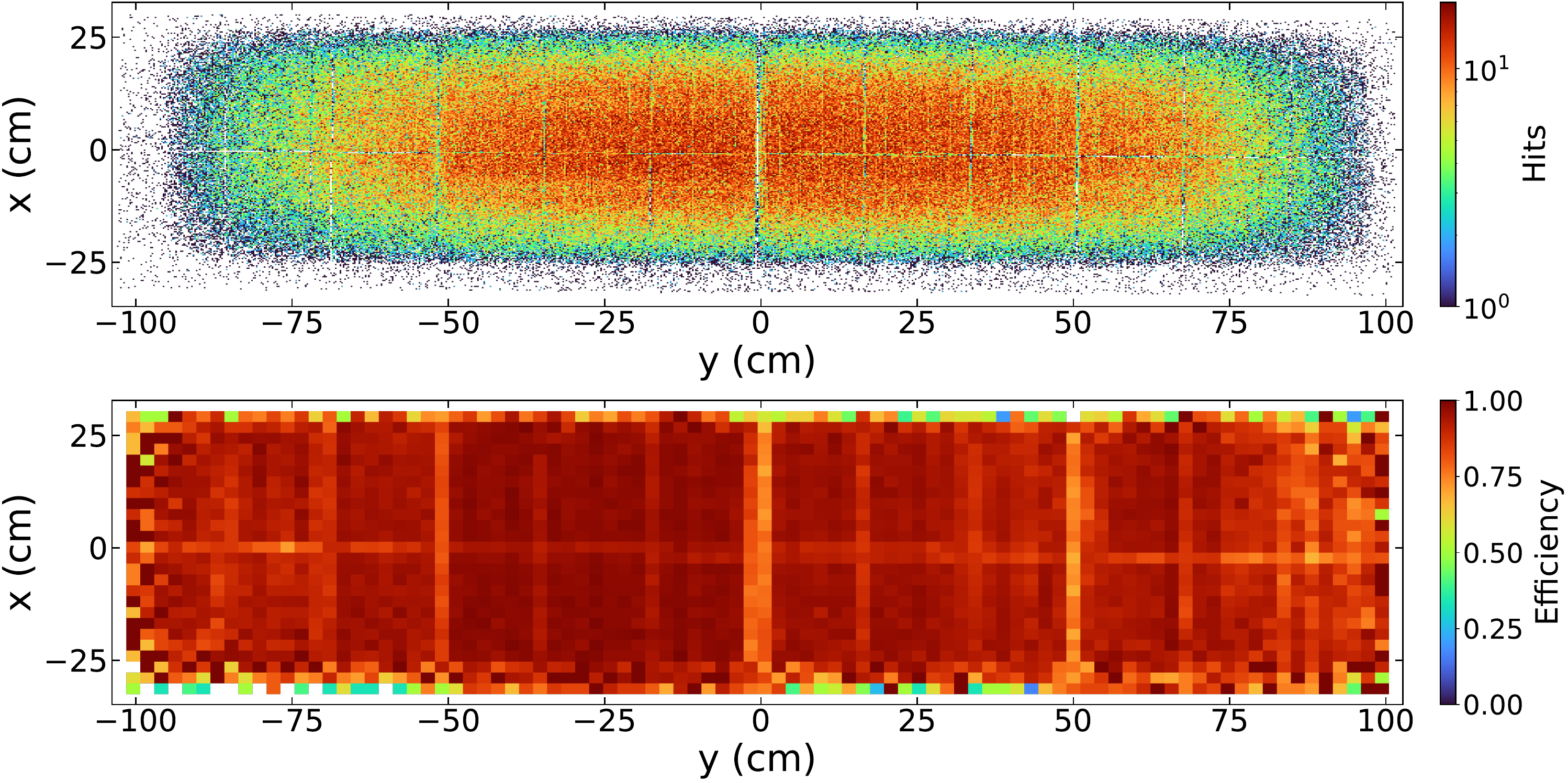}
    \caption{Layer 4}
    \label{fig:eff_L3}
  \end{subfigure}

  \vspace{1.0em}

  \begin{subfigure}[t]{0.49\textwidth}
    \centering
    \includegraphics[width=\linewidth]{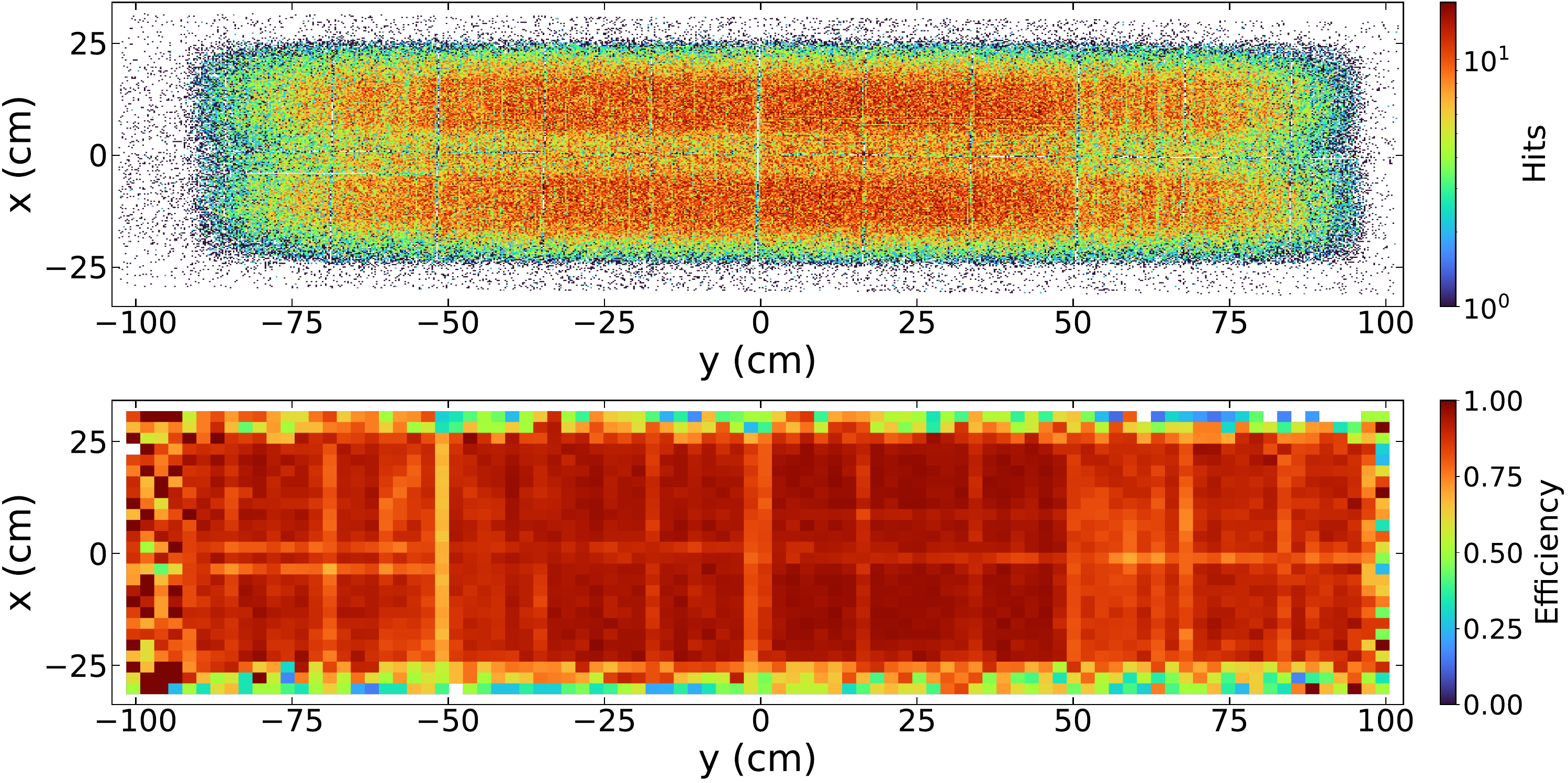}
    \caption{Layer 5}
    \label{fig:eff_L4}
  \end{subfigure}

  \caption{Two dimensional hit (top) and efficiency (bottom) maps for GEM layers 1--5.}
  \label{fig:eff_221}
\end{figure*}

\subsection{Active Area Uniformity and Efficiency}
The top panels of \cref{fig:eff_L0,fig:eff_L1,fig:eff_L2,fig:eff_L3,fig:eff_L4} show  2D hit-maps of the hits that participated in the reconstruction of tracks from cosmic rays. This is a set of 5 XY GEM layers, vertically stacked on the test stand. A minimum of 3 hits per track was required for this analysis. Finer physical details of individual GEM modules are evident, such as spacer grids within the active area and small gaps between module boundaries. The reconstructed 2D hit distributions narrow progressively with increasing layer number. This effect arises because the top trigger scintillator bars were narrower than the GEM active area, restricting acceptance near the detector edges and in the gap between adjacent bars. Some features due to inefficient trigger scintillators were observed, especially in the area between -100 cm and 0 cm on the horizontal axis ($y$), and -20 cm and 0 cm on the vertical axis ($x$) of Layer 1.

The bottom panels of \cref{fig:eff_L0,fig:eff_L1,fig:eff_L2,fig:eff_L3,fig:eff_L4} show the \textit{inclusive} track based efficiency. The GEM module/layer under question is not excluded in the reconstruction of the tracks used to evaluate the efficiency, and thus the efficiency reported is referred to as the inclusive track based efficiency.

The inclusive method is biased by the fact that the hits from the GEM module/layer in question are not excluded in the fitted tracks used to measure the efficiency. This is in contrast to \textit{exclusive} measures of efficiency in which the module/layer whose efficiency is being measured is excluded from participating in the track formation. However, tracks used for this inclusive efficiency analysis are required to have at least 3 hits in modules/layers \textit{other} than the module/layer of which the efficiency is measured, somewhat mitigating the ``survivor-ship bias'' inherent in the inclusive track-based efficiency. It is also worth noting that as a result of the low occupancy and background conditions of cosmic data taking, the reconstructed tracks are real with high probability, which is crucial to the accuracy of a track-based efficiency analysis. 

The 2D efficiency maps shown in the bottom panels of  \cref{fig:eff_L0,fig:eff_L1,fig:eff_L2,fig:eff_L3,fig:eff_L4} are generated as follows. First, a coarsely binned 2D histogram is generated for each layer, called a ``should hit'' histogram. These ``should hit'' histograms are filled by projecting tracks to each layer and incrementing the bin content whenever a track crosses a given bin in the 2D histogram, regardless of whether it includes a hit in that layer. Then, an identical set of 2D ``did hit'' histograms is generated for each layer. These ``did hit'' histograms are populated by looping through all the hits that participated in reconstructing the above set of tracks (that were used to generate the ``should hit'' histograms) and incrementing the histogram bin content of the bin that the track intercepts. The efficiency in each bin is then defined as the bin content ratio of the histograms ``did hit'' to ``should hit''.

The loss in efficiency in the top right corner of the third module from the left of layer 1 (the bottom panel of \cref{fig:eff_L0}) is due to a compromised gas flow to that region. The imperfect alignment of GEM modules within the active area of the layer is evident in the narrow vertical stripes of low reported efficiency near the module boundaries, especially in layers 1, 2, and 3. These vertical stripes are caused by small gaps between the active areas of the GEM modules, and the degree of efficiency reduction reported reflects the size of the gap relative to the width of the histogram bin. 

\begin{figure}[!]
    \centering
    \includegraphics[width=\linewidth]{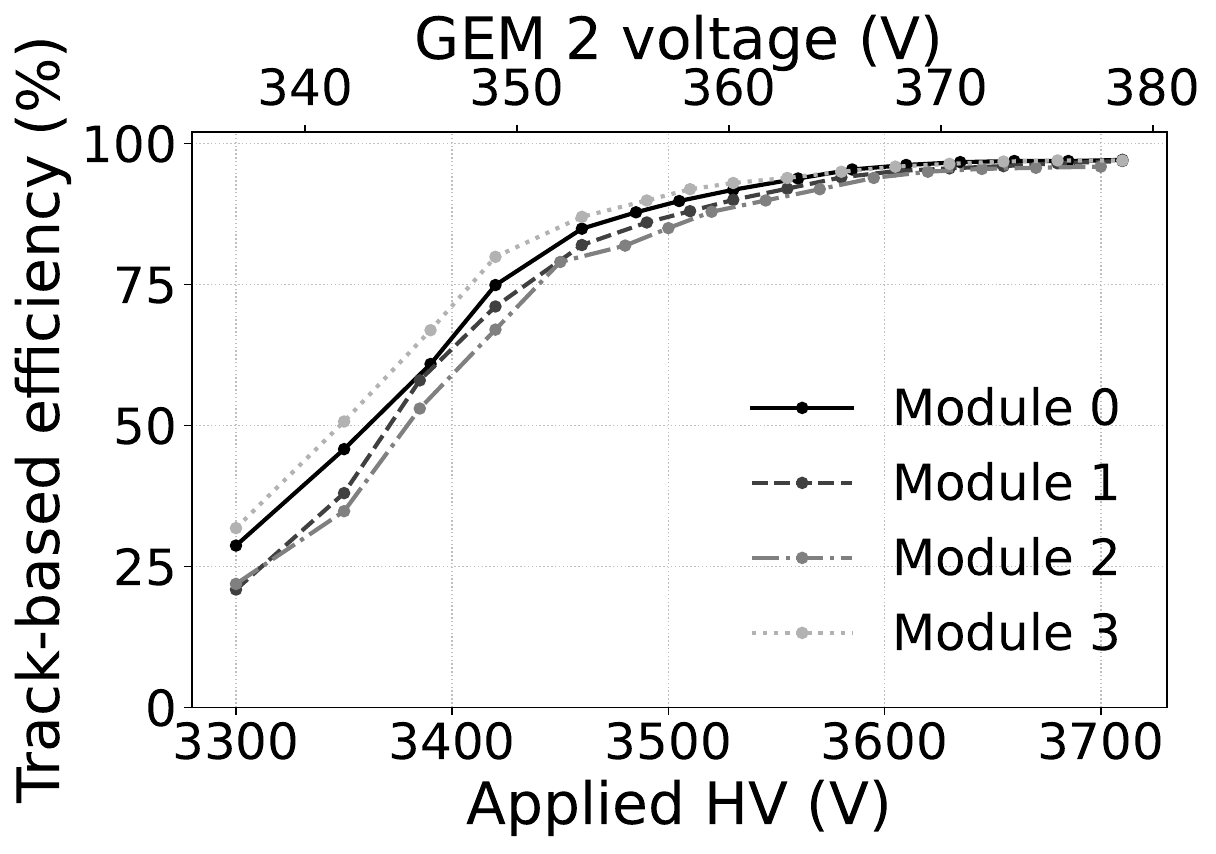}
    \caption{Track-based efficiency as a function of the applied voltage for four XY GEM modules.}
    \label{fig:Track-based efficiency Layer 6 all modules.}
\end{figure}

The studies of 2D hit and efficiency maps were vital to understand the individual GEM module and layer performances. The GEM modules with lower efficiency and uniformity were identified during this process and were moved to the outer edges of the tracking layers, away from the central spectrometer acceptance.

\subsection{Efficiency Plateau Studies}~\label{subscetion:Plateau_Studies}
High-efficiency operation of individual GEM modules while ensuring reliability is an important factor to maximize experiment productivity. The GEM detection efficiency for the Minimum Ionizing Particles (MIP), with respect to the applied high-voltage, follows a sigmoid-like function. The quantities of primary interest are the plateau efficiency and the minimum high voltage required to reach the efficiency plateau.

\Cref{fig:Track-based efficiency Layer 6 all modules.} shows the track based efficiency versus applied high-voltage for four XY GEM modules within a single GEM layer. All four GEM modules have plateau efficiencies greater than 95$\%$. However, the minimum high voltage required to reach the plateau efficiency is GEM module dependent. In certain instances, operating the GEM module at voltages higher than necessary to reach the plateau increases the probability of discharges, which could damage the GEM foils. Therefore, high voltage plateau characterizations were conducted for every GEM module and operating voltages were conservatively chosen by selecting a value corresponding to a reduction from the plateau efficiency by no more than 2\%.

\section{Summary}\label{Summary} A total of 56 large-area triple GEM modules were assembled for the SBS nucleon form factor program in experimental Hall A at the Thomas Jefferson National Accelerator Facility. These modules were characterized by cosmic-ray testing prior to installation as part of different spectrometer configurations across six different experiments. Significant developments in both GEM hardware and software were made to accommodate the high-luminosity operation of the GEM detectors at data throughput of tens of gigabits per second. Extensive cosmic-ray testing and in-beam commissioning of the GEM modules revealed several areas of necessary improvement. To mitigate the effects of CM fluctuations in the APV25 readout electronics, individual RF shielding was added to the GEM modules. The stability of several GEM modules was improved by a 15$\%$ reduction in transfer field strengths. Under high rates, GEM detector gains were observed to be limited by the initial passive resistive high-voltage dividers. To resolve these gain drops, a more sophisticated individual-channel high-voltage scheme was developed and successfully deployed. Cosmic-ray studies revealed excellent active area uniformity and track-based efficiency in excess of 90$\%$ for the SBS GEM modules. The timing and spatial resolutions were consistent with the expected performance of triple-GEM detectors, within the evaluation limitations of the cosmic-ray test setup.

\section{Acknowledgment}\label{Acknowledgement}The authors acknowledge and extend their gratitude for the support of the SBS Collaboration, the technical staff of Halls A and C of JLab, the JLab electronics group,in particular Mahlon (Chuck) Long, Jack Segal, Catherine Feldman, and David Armstrong. The authors also acknowledge the contributions of the INFN and ISS technical staff (Roma, Catania) in the early stages of commissioning. This material is based upon the work supported by the United States Department of Energy, Office of Science, Office of Nuclear Physics, under contract No.\ 89243126CSC000213. This work was also supported in part by the United States Department of Energy, Office of Science, awards DE-FG02-03ER41240, DE-SC0021200, and DE-SC0013941 and by National Science Foundation awards PHY-1714792, PHY-2012738, and PHY-2412825.
The authors acknowledge support from the United States Department of Energy, Office of Science Graduate Student Research Fellowship Award, and from the JSA Graduate Fellowship, administered by Jefferson Science Associates and the Southeastern Universities Research Association.
INFN fellows acknowledge support by the EU Horizon 2020 Research and Innovation Program under the Marie Sklodowska-Curie Grant Agreement No.\ 101003460.

\end{document}